\documentclass[a4paper,11pt]{article}
\usepackage{jheppub} 
\usepackage{lineno}

\usepackage{braket}
\usepackage{booktabs}

\arxivnumber{} 

\title{\boldmath Topological invariant for holographic Weyl-Nodal line coexisting semimetal}

\author[a]{Xiantong Chen}
\author[a,b]{Xuanting Ji}
\author[a,c]{Ya-Wen Sun}
\affiliation[a]{
    School of Physical Sciences,\\ 
    University of Chinese Academy of Sciences, Beijing 100049, China
}
\affiliation[b]{
    Department of Applied Physics, College of Science, \\
    China Agricultural University, Beijing 100083, China
}
\affiliation[c]{
    Kavli Institute for Theoretical Sciences, \\
    University of Chinese Academy of Sciences, 
    Beijing 100049, China
}

\emailAdd{{chenxiantong23@mails.ucas.ac.cn}, {jixuanting@cau.edu.cn}, {yawen.sun@ucas.ac.cn}}

\abstract{
    The presence of a topological phase in a topological many-body system can be distinguished through the analysis of topological invariants. In the present study, the topological invariants for the strongly coupled holographic semimetals have been systematically computed, especially focusing on the holographic Weyl-Nodal line coexisting semimetal. The topological invariants that we calculate include the Weyl charge, the topological charges for a nodal ring $\zeta_0$, $\zeta_1$, $\zeta_2$ and an additional mirror symmetry protected  topological invariant, $\widetilde{\zeta}_{2}$, that we herein introduce. In addition, the effective band structures and topological invariants in the critical phases of holographic semimetals are investigated, including the case of Weyl, nodal line and Weyl-Nodal line coexisting semimetals. The findings indicate the presence of notable and unique features inherent to strongly coupled topological semimetals, including band-crossing ordering interchange and multi Fermi surfaces, which provide a valuable platform for experimental investigations of strongly coupled semimetals in condensed matter physics.
}

\begin{document}
\maketitle
\flushbottom

\section{Introduction}
    \label{Section1}
    Topological semimetals are a unique class of quantum materials featuring protected band-crossing points in momentum space\cite{TopologicalSemimetalswan, TopologicalPhaseMatterWitten:2015aoa}. These materials have attracted significant attention due to their stable topological properties and distinctive physical phenomena, including surface Fermi arc\cite{ConductiveSurfaceStateLiu_2014}, drumhead Fermi surface\cite{BulkEdgeCorrespondenceTKNN}, anomalous Hall conductivity\cite{QuantilizedHallConductanceChang_2013, QuantumHallEffectThouless1, QuantumHallEffectThouless2}, negative magneto-resistivity\cite{NegativeMagnetoResistivityHuang_2015}, etc. Unlike accidental band crossings in conventional materials, the crossings in topological semimetals remain stable against perturbations. In order to reliably identify such protected crossings, and distinguish topological semimetals from ordinary materials, the traditional band theory was combined with the topology theory and the so-called topological invariants can be defined for topological states. These can then be utilized to distinguish different types of topological states \cite{TopologicalNumberkm1, TopologicalNumbermoore, TopologicalNumberkm2}.

    Topological invariants are quantities that remain unchanged under topological transformations. They are independent of the choice of gauge (e.g., connection) and are solely determined by the global topological properties of the system. In quantum systems, these invariants exhibit robustness against perturbations, making them ideal tools for characterizing topological semimetals.

    In weakly coupled topological semimetals, topological invariants are typically calculated by integrating the Berry connection or Berry curvature, with the specific form depending on both the physical system and the construction of the invariants. Taking Weyl semimetals and nodal line semimetals as examples: although both possess topologically protected band structures, their distinct band geometries (discrete points vs. continuous lines) lead to fundamentally different topological properties.

    Weyl semimetals represent a class of topological semimetals characterized by the emergence of paired Weyl nodes in momentum space, resulting from broken TP symmetry. In these systems, the topological charge - known as the Weyl charge - serves as the fundamental topological invariant. This quantity can be directly computed by integrating the Berry curvature over a closed surface enclosing a Weyl point. A non-zero integration value provides definitive evidence of a topologically protected Weyl node within the enclosed region\cite{ReviewWeylSemimetalsArmitage}.

    Nodal line semimetals are topological systems where conduction and valence bands cross along closed loops (nodal lines) in momentum space. Protected by TP symmetries, these systems exhibit strictly vanishing topological charges via conventional Berry curvature integration, necessitating higher-order invariants $\zeta_1$ and $\zeta_2$: (1) $\zeta_1$ invariant: Obtained by integrating the Berry connection along a loop linked with the nodal line, where non-zero values certify topological protection; (2) $\zeta_2$ invariant: Computed via nested Wilson loops over a surface enclosing the nodal line, with non-zero values indicating that the nodal line can be continuously deformed to a point but will inevitably reopen, reflecting its topological stability\cite{TopologicalNodalLineSemimetalsFang_2016}.

    Moreover, since a nodal ring always lies within a specific plane, in systems possessing mirror symmetry with respect to that plane, this symmetry plays a crucial role in governing the ring's topological properties. 
    In systems protected by mirror symmetry, topological invariants can be defined  high-symmetry points, reflecting topological properties under perturbations that obey the mirror symmetry. Then two additional topological invariants playing  similar roles as the $\zeta_1$ and $\zeta_2$ invariants could be defined, which we denote as $\zeta_0$ and  $\widetilde{\zeta}_2$. $\zeta_0$ is calculated by comparing the eigenvalues under the mirror operator at points inside and outside the nodal ring, while  $\widetilde{\zeta}_2$ is obtained by comparing the average phases for the eigenvalues of the Wilson loops along closed paths interior and exterior to the nodal ring within the mirror plane.

    In coexisting Weyl-nodal line semimetals, partial breaking of TP symmetries allows the simultaneous presence of Weyl nodes and nodal lines, necessitating a combined characterization via Weyl charges, $\zeta_1$, and $\zeta_2$ invariants. The interplay between these topological states generate a rich phase diagram, including: Weyl-Nodal, Weyl-Critical, Weyl-Gap, Critical-Nodal, Critical-Critical, Critical-Gap, Gap-Nodal, Gap-Critical, Gap-Gap phases. Transitions between these phases (e.g. Weyl-Critical $\to$ Gap-Nodal) give rise to interesting quantum criticality and transport phenomena.

    Unlike weakly coupled systems where topological invariants can be obtained via direct integration, strongly correlated systems lack a complete Hamiltonian description. The AdS/CFT principle resolves this through the topological Hamiltonian approach—constructing an effective Hamiltonian via holographic models and probe fermions to extract topological invariants\cite{GKPWwitten, GKPWgubser, HolographicTopologicalInvariantLiu_2018, HolographicSemimetalsLandsteiner:2019kxb, HolographicWeylSemimetalsLandsteiner:2015lsa, HolographicWeylSemimetalsLandsteiner:2016stv, HolographicWeylSemimetalsPlantz_2018}. Our prior work successfully applied this method to:
    (1) Weyl charges in holographic Weyl semimetals\cite{HolographicWeylZ2Ji_2021, HolographicWeylZ2chen2025},
    (2) Weyl and $\mathbb{Z}_2$ charges in holographic Weyl-$\mathbb{Z}_2$ semimetals\cite{HolographicWeylZ2Ji_2021, HolographicWeylZ2chen2025},
    (3) $\zeta_1$ invariants in holographic nodal line semimetals\cite{HolographicNodalLineSemimetalsLiu_2021, TopologicalNodalLineSemimetalsFang_2016}.

    However, three critical questions remain open. The first one is about the holographic coexisting Weyl-nodal line semimetal, whose topological invariants are still unexplored. This would provide a more profound comprehension of the phase structure of the system.
    The second question is that, to date, only the invariant $\zeta_1$ in the holographic nodal line systems has been computed; the calculation of $\zeta_2$ has yet to be undertaken. Moreover, as previously stated, the mirror symmetry will introduce two additional topological invariants: $\zeta_0$ and $\widetilde{\zeta}_2$, which have not been calculated yet.
    The third question is the calculation of topological invariants and effective band structures for critical phases of holographic semimetals. For holographic semimetals, the critical phase corresponds to a bulk Lifshitz geometry, and it marks the transition between trivial and topologically nontrivial phases. Weak-coupling studies show that at the critical phase, Weyl nodes merge into Dirac points (the Weyl charge is 0) and nodal lines collapse to Dirac points ($\zeta_1$=0). Verifying this behavior in holography is essential for the study of strongly coupled topological semimetals.

    This work thus aims to compute all topological invariants (Weyl/$\zeta_0$/$\zeta_1$/$\zeta_2$/$\widetilde{\zeta}_{2}$) for Weyl-nodal line coexisting semimetal phases and systematically study critical-phase properties across Weyl/nodal-line/coexisting systems. This article is structured as follows. Section \ref{Section2} begins by reviewing the coexisting Weyl-nodal line semimetals both in the weak coupled limit and in holography. In section \ref{Section3}, we first present the definition and calculation procedures for topological invariants (Weyl/$\zeta_0$/$\zeta_1$/$\zeta_2$/$\widetilde{\zeta}_{2}$) and the results in weak coupling field theory models for coexisting semimetals. Then we calculate the topological invariants for the Weyl-Nodal coexisting phase of holographic coexisting semimetals. We explicitly present the bulk action of probe fermions and employ the topological Hamiltonian method to calculate topological invariants for the Weyl-Nodal coexisting phase. Through the effective band structure along $k_x$ and $k_z$ axes, we report the first observation of band crossing ordering interchange phenomena in the holographic nodal line systems-a feature previously found in holographic Weyl semimetals \cite{HolographicWeylZ2chen2025}. Section \ref{Section4} focuses on the topological properties of critical phases: using corresponding IR geometries, we compute topological invariants for holographic Weyl semimetals, nodal-line semimetals, and the Weyl-Critical phase for coexisting semimetals, supplemented by characteristic effective band structures. Notably, despite both being critical phases,  holographic Weyl and nodal-line systems exhibit distinct critical phase band structure features. The calculated topological invariants demonstrate remarkable consistency with weak-coupling results, validating the reliability of our holographic framework. It should be noted, however, that due to some numerical difficulties encountered in the Critical-Nodal phase, and also because its behavior closely parallels that of the Weyl-Critical phase and the critical phase in the Weyl semimetal, we do not perform the explicit computation of topological invariants for the Critical-Nodal phase without affecting the overall conclusions. Finally, Section \ref{Section5} summarizes key findings and discusses open questions.

\section{Review of the coexisting Weyl-nodal line semimetal}
    \label{Section2}
    Weyl and nodal line semimetals are two distinctive members of the topological semimetal family, characterized by distinct band structures and topology properties. It can thus be concluded that the topological phase transition processes involved in Weyl and nodal line semimetals are quite different from each other, and can be characterized by the change of their corresponding distinct topological invariant. The topological phase transitions in pure Weyl semimetal and nodal line semimetal have already been the subject of study; however, it is both interesting and natural to pose the following question: what would be the phase structure in the system where these two topological semimetal states could coexist? In previous works, two theoretical models were constructed – a field theory model\cite{TopologicalPhaseTransitionJi:2023rua} and a holographic model\cite{HolographicWeylNodalChu_2024} – to demonstrate the coexistence of a Weyl and a nodal line semimetal in the weak and strong coupling limit, respectively. 
    
    This section systematically reviews previous works on the coexisting Weyl-nodal line semimetal system to establish the foundation for subsequent analysis. This review contains the following two parts, (1) Weak coupling field theoretical models. In this part we will present the field theoretical models for Weyl semimetal-a TP symmetry broken model with isolated band crossings and for nodal line semimetal-a TP symmetry protected model with ring shaped degeneracies. Then the coexisting Weyl-nodal line semimetal can be constructed by partially breaking TP symmetry. (2)A strongly coupled holographic model. In this part the holographic model for strong coupling Weyl semimetal and nodal line semimetal will be reviewed. Then based on these two models the holographic model for coexisting Weyl-nodal line model will be presented. 
    \subsection{Effective field theoretical model for the coexisting Weyl-nodal line semimetal}
        In this part we will review the weak coupling field theoretical model\cite{TopologicalSemimetalsWeakCouplingModelGrushin:2012mt} for topological semimetals. As discussed in previous work, the Lagrangian for a weak coupling Weyl semimetal is
        \begin{equation}
            \label{PureWeylLagrangian}
            \mathcal{L}_{\text{Weyl}}=i\Bar{\Psi}(\Gamma^\mu\partial_\mu-i\Gamma^5\Gamma^\mu A_\mu+M)\Psi,
        \end{equation}
        where $i\Gamma^5\Gamma^\mu$ is the so-called ``Lorentz breaking" term breaking the $TP$ symmetry to form Weyl nodes, $M$ is the mass term and $\Psi$ is a four components spinor. To ensure the location of the Weyl nodes we can perform Legendre and Fourier transformation for \eqref{PureWeylLagrangian} to get the Hamiltonian in the momentum space
        \begin{equation}
            \label{WeylHamiltonian}
            H_{\text{Weyl}}=i\Gamma^0(-ik_i\Gamma^i+i\Gamma^5\Gamma^\mu A_\mu-M_1).
        \end{equation}
        
        Without loss of generality we take the ansatz $A_\mu =A_z {\delta_\mu}^z$. There are a pair of Weyl nodes at $k_z=\pm\sqrt{{A_z}^2-M^2}$ from the spectrum.
        
        Similarly the Lagrangian for a weak coupling nodal line semimetal is
        \begin{equation}
            \label{PureNodalLagrangian}
            \mathcal{L}_{\text{nodal}}=i\Bar{\Psi}(\Gamma^\mu\partial_\mu-\Gamma^{\mu\nu}b_{\mu\nu}-\Gamma^{\mu\nu}\Gamma^5{b_{\mu\nu}}^5+M)\Psi,
        \end{equation}
        where $b_{\mu\nu}$ is an antisymmetric real two-form field, and the term $\Gamma^{\mu\nu}b_{\mu\nu}$ contributes to the formation of the nodal line semimetal, $M$ is the mass term and $\Psi$ is a four component spinor.  According to the duality relation between the anti-symmetric tensor $\Bar{\Psi}\Gamma^{\mu\nu}\Gamma^5\Psi=-\frac{i}{2}{\epsilon_{\alpha\beta}}^{\mu\nu}\Bar{\Psi}\Gamma^{\alpha\beta}\Psi$ in the four-dimensional Minkowski spacetime, a pure imaginary dual part of the $b_{\mu\nu}$ needs to be introduced, which is ${b_{\mu\nu}}^5$. Using the same method we get the Hamiltonian for nodal line semimetal in the momentum space
        \begin{equation}
            \label{NodalHamiltonian}
            H_{\text{nodal}}=i\Gamma^0(-ik_i\Gamma^i+\Gamma^{\mu\nu}b_{\mu\nu}+\Gamma^{\mu\nu}\Gamma^5 b_{\mu\nu}^5-M_2).
        \end{equation}
        
        Without loss of generality we fix all the non-zero components for $b_{\mu\nu}$ and $b_{\mu\nu}^5$ as $b_{xy}=-b_{yx}=B_{xy}$ and $b_{tz}^5=-b_{zt}^5=iB_{xy}$. There will be a nodal ring at the radius $\sqrt{16 {B_{xy}}^2-M^2}$.
         
        The nodal line is protected by the $TP$ symmetry, while the existence of Weyl nodes requires the breaking of the $TP$ symmetry. Therefore, we define the Lagrangian for the coexisting Weyl-nodal line topological semimetal as\cite{TopologicalPhaseTransitionJi:2023rua}
        \begin{equation}
            \label{WeylNodalCoexistLagrangian}
            \mathcal{L}=i\Bar{\Psi}[(\Gamma^\mu\partial_\mu-i\Gamma^5\Gamma^\mu A_\mu+M_1)\oplus(\Gamma^\mu\partial_\mu-\Gamma^{\mu\nu}b_{\mu\nu}-\Gamma^{\mu\nu}\Gamma^5b_{\mu\nu}^5+M_2)]\Psi,
        \end{equation}
        where $\Psi$ is an eight-component spinor. To find the Weyl nodes and nodal lines in the band structure , we perform Legendre transformation on the Lagrangian \eqref{WeylNodalCoexistLagrangian} and perform Fourier transformation to get the Hamiltonian in the momentum space
        \begin{equation}
            \label{WeylNodalCoexistHamiltonian}
            H=[i\Gamma^0(-ik_i\Gamma^i+i\Gamma^5\Gamma^\mu A_\mu-M_1)]\oplus[i\Gamma^0(-ik_i\Gamma^i+\Gamma^{\mu\nu}b_{\mu\nu}+\Gamma^{\mu\nu}\Gamma^5b_{\mu\nu}^5-M_2)].
        \end{equation}
        
        Without loss of generality we can take the ansatz $A_\mu=A_z{\delta_\mu}^z$ and $b_{xy}=-b_{yx}=B_{xy},b_{tz}^5=-b_{zt}^5=iB_{xy}$, then there will be a pair of Weyl nodes along the $k_z$ axis at $k_z=\pm\sqrt{{A_z}^2-{M_1}^2}$ and a nodal line at $k_z=0$ plane with radius $\sqrt{16{B_{xy}}^2-{M_2}^2}$ in the momentum space.
        
        Two dimensionless parameters, $M_1/b$ and $M_2/c$ (where $b\equiv A_z$ and $c\equiv B_{xy}$), serve as crucial indicators for phase transitions in coexisting Weyl-nodal line semimetal. Their values directly determine the phase of the system. In this weak coupling field theoretical model, for the Weyl sector $(M_1/b)$ we have: (1) $M_1/b<1$,  separated Weyl nodes; (2) $M_1/b = 1$, Weyl nodes merged  to form a Dirac point; (3) $M_1/b > 1$, fully gapped spectrum. Similarly, for the nodal line sector $(M_2/c)$, we have: (1) $M_2/c < 4$, stable nodal ring; (2) $M_2/c = 4$, ring collapse to a Dirac point; (3) $M_2/c > 4$, fully gapped spectrum. In the field theoretical model for Weyl-nodal line coexisting semimetal, the parameters $M_1/b,M_2/c$ denote nine distinct phases, systematically categorized as: Weyl-Nodal, Weyl-Critical, Weyl-Gap, Critical-Nodal, Critical-Critical, Critical-Gap, Gap-Nodal, Gap-Critical, and Gap-Gap phases. The complete phase diagram is presented in Fig. \ref{CoexistPhaseDiagram}, while two critical lines $M_1/b =1$ and $M_2/c=4$ indicate topological phase transitions.
    \subsection{Holographic model for the coexisting Weyl-Nodal line semimetal}
        Let us first review the holographic models for the Weyl semimetal and nodal line semimetal, which provide the basic elements for building the holographic model for the Weyl-nodal line coexisting semimetal. The action of the holographic model for Weyl semimetals is \cite{HolographicTopologicalInvariantLiu_2018}
        \begin{equation}
            \label{WeylHolographicModel}
            \begin{aligned}
                S_{\text{Weyl}}&=\int d^5x\sqrt{-\det g}\left[\frac{1}{2\kappa^2}\left(R+\frac{12}{L^2}-\frac{{F_V}^2}{4}-\frac{{F_A}^2}{4}\right)\right.\\
                &+\left.\frac{\alpha}{3}\epsilon^{abcde}A_a\left(3{F}_{Vbc}{F}_{Vde}+{F}_{Abc}{F}_{Ade}\right)-(D_a\Phi)^* D^a\Phi-m^2|\Phi|^2-\frac{\lambda}{2}|\Phi|^4\right],
            \end{aligned}
        \end{equation}
        where $g$ is the metric for 5-dimensional AdS space time and $R$ is the corresponding Ricci curvature, $V$ is a vector gauge field and $A$ is an axial gauge field with $F_V,F_A$ their corresponding field strengths. $\Phi$ is a scalar field providing the effective mass term for the holographic Weyl semimetal. $\kappa^2$ is a 5-dimensional gravitational constant, $L$ is the AdS radius. $\alpha$ is the coupling constant for Chern-Simons term, $m$ and $\lambda$ are the coefficients of the potential for $\Phi$. $D_a=\partial_a-iqA_a$ is the covariant derivative for $\Phi$ and $q$ is the corresponding coupling constant.
        
        Similarly, the holographic model for the nodal line semimetal is \cite{HolographicTopologicalInvariantLiu_2018}
        \begin{equation}
            \label{NodalHolographicModel}
            \begin{aligned}
                S_{\text{nodal}}&=\int d^5x\sqrt{-\det g}\left[\frac{1}{2\kappa^2}\left(R+\frac{12}{L^2}-\frac{{F_V}^2}{4}-\frac{{F_A}^2}{4}\right)\right.\\
                &+\left.\frac{\alpha}{3}\epsilon^{abcde}A_a\left(3{F}_{Vbc}{F}_{Vde}+{F}_{Abc}{F}_{Ade}\right)-(D_a\Phi)^* D^a\Phi-{m}^2|\Phi|^2-\frac{\lambda}{2}|\Phi|^4\right.\\
                &-\left.{m_B}^2 {B_{ab}}^*B^{ab}-\lambda|\Phi|^2{B_{ab}}^*B^{ab}-\frac{1}{6\eta}\epsilon^{abcde}\left(iB_{ab}{H_{cde}}^*-i{B_{ab}}^*H_{cde}\right) \right],
            \end{aligned}
        \end{equation}
        where $B$ is a 2-form field and $H_{abc}=\partial_a B_{bc}+\partial_b B_{ca}+\partial_c B_{ab}-iq{A}_aB_{bc}-iq{A}_bB_{ca}-iq{A}_cB_{ab}$. $m_B$ and $\lambda_B$ are the coefficients of the potential for $B$. $\eta$ is a coupling constant. 
        
        Based on the holographic model for Weyl and nodal line semimetals and the dictionary of the gauge/gravity duality, a holographic model to realize a state where the Weyl semimetal and the nodal line semimetal coexist can be obtained as follows\cite{HolographicWeylNodalChu_2024}
        \begin{equation}
            \label{CoexistHolographicModel}
            \begin{aligned}
                S&=\int d^5x\sqrt{-\det g}\left[\frac{1}{2\kappa^2}\left(R+\frac{12}{L^2}\right)-\frac{{F_V}^2}{4}-\frac{{\hat{F}_V}^2}{4}-\frac{{F_A}^2}{4}-\frac{{\hat{F}_A}^2}{4}\right.\\
                &+\frac{\alpha}{3}\epsilon^{abcde}A_a\left(3{F}_{Vbc}{F}_{Vde}+3{{\hat{F}}_{Vbc}}{\hat{F}}_{Vde}+{F}_{Abc}{F}_{Ade}+{{\hat{F}}_{Abc}}{\hat{F}}_{Ade}\right)\\
                &+\frac{2\beta}{3}\epsilon^{abcde}\hat{A}_a\left(3{\hat{F}}_{Vbc}{F}_{Vde}+{{\hat{F}}_{Abc}}{F}_{Ade}\right)-\frac{1}{6\eta}\epsilon^{abcde}\left(iB_{ab}{H_{cde}}^*-i{B_{ab}}^*H_{cde}\right)\\
                &-(D_a\Phi_1)^*D^a\Phi_1-{m_1}^2|\Phi_1|^2-(\hat{D}_a\Phi_2)^*\hat{D}^a\Phi_2-{m_2}^2|\Phi_2|^2-{m_3}^2{B_{ab}}^*B^{ab}\\
                &-\left.\frac{\lambda_1}{4}(|\Phi_1|^4+|\Phi_2|^4)-\lambda|\Phi_2|^2{B_{ab}}^*B^{ab}\right],
            \end{aligned}
        \end{equation}
        where $A,\hat{A}$ are introduced as two axial gauge field, $V,\hat{V}$ are introduced as two vector gauge field, and $F_A=dA,~\hat{F}_A=d\hat{A},~F_V=dV,~\hat{F}_V=d\hat{V}$ are corresponding field strength. This is a combination of the holographic model for pure Weyl semimetal and pure nodal line semimetal, where $A$, $V$ are inherited from the Weyl holographic model \eqref{WeylHolographicModel} and $\hat{A}$, $\hat{V}$ are inherited from the nodal line holographic model \eqref{NodalHolographicModel}. 
        Note that in order to partially break the $TP$ symmetry, thereby ensuring the existence of Weyl nodes, and partially preserve the mirror symmetry, thus ensuring the existence of a nodal line semimetal state, it is necessary to employ two sets of $U(1)$ gauge fields, $V$ and $\hat{V}$, and two sets of axial $U(1)$ gauge fields $A$ and $\hat{A}$. $B$ is the complex 2-form field which plays the same role as in the pure nodal line semimetal and $H$ is defined through $\hat{A}$ and $B$ as $H_{abc}=\partial_a B_{bc}+\partial_b B_{ca}+\partial_c B_{ab}-iq_3\hat{A}_aB_{bc}-iq_3\hat{A}_bB_{ca}-iq_3\hat{A}_cB_{ab}$. $\Phi_1,\Phi_2$ are two scalar fields which provide the effective mass terms for the fermions in the holographic coexisting semimetal, and the covariant derivatives for $\Phi_1,\Phi_2$ are defined as $D_a\Phi_1=(\partial_a-iq_1 A_a)\Phi_1$, $\hat{D}_a\Phi_2=(\partial_a-iq_2 \hat{A}_a)\Phi_2$. $g$ is the metric for the 5-dimensional AdS spacetime and $R$ is the corresponding scalar curvature. $\kappa^2$ is the 5-dimensional gravitational constant and $L$ is the AdS radius.
        
        Without loss of generality we set  $2\kappa^2=L=1$, and fix the coupling constants for the Chern-Simons terms $\alpha=\beta=1,~\eta=2$, the mass term ${m_1}^2={m_2}^2=-3, m_3=1$ and the coupling constants for the scalar fields $\lambda=1,~\lambda_1=\frac{1}{10}$, the coupling constants for axial gauge fields $q_1=\frac{1}{2},~ q_2=\frac{3}{2},~ q_3=\frac{1}{4}$.
        
        For simplicity, we can take the following ansatz for the fields employed in the holographic model for the Weyl-nodal line coexisting semimetal \eqref{CoexistHolographicModel} at zero temperature
        \begin{equation}
            \label{BackGroundZeroTemperatureAnsatz}
            \begin{aligned}
                g&=-u dt^2+f dx^2+fdy^2+hdz^2+\frac{dr^2}{u},\\
                A&=A_zdz,~B=\frac{1}{2}\left(B_{xy}dx\wedge dy+iB_{tz}dt\wedge dz\right),\\
                \Phi_1&=\phi_1,~\Phi_2=\phi_2,
            \end{aligned}
        \end{equation}
        where all the nonzero components $u,f,h,A_z,B_{xy},B_{tz},\phi_1,\phi_2$ are real functions of the radial coordinate $r$.
        
        Under the ansatz \eqref{BackGroundZeroTemperatureAnsatz}, the boundary asymptotic behaviors of each field are explicitly given below
        \begin{equation}
            \label{BackGroundBoundaryBehaviour}
            \begin{aligned}
                \lim_{r\to\infty}A_z&=b,~\lim_{r\to\infty}\frac{B_{xy}}{r}=\lim_{r\to\infty}\frac{B_{tz}}{r}=c,\\
                \lim_{r\to\infty}r\phi_1&=M_1,~\lim_{r\to\infty}r\phi_2=M_2.
            \end{aligned}
        \end{equation}
        
\begin{figure}[htbp]
            \begin{minipage}{0.49\linewidth}
                \vspace{3pt}
                \centerline{\includegraphics[width=\textwidth, height=\textwidth]{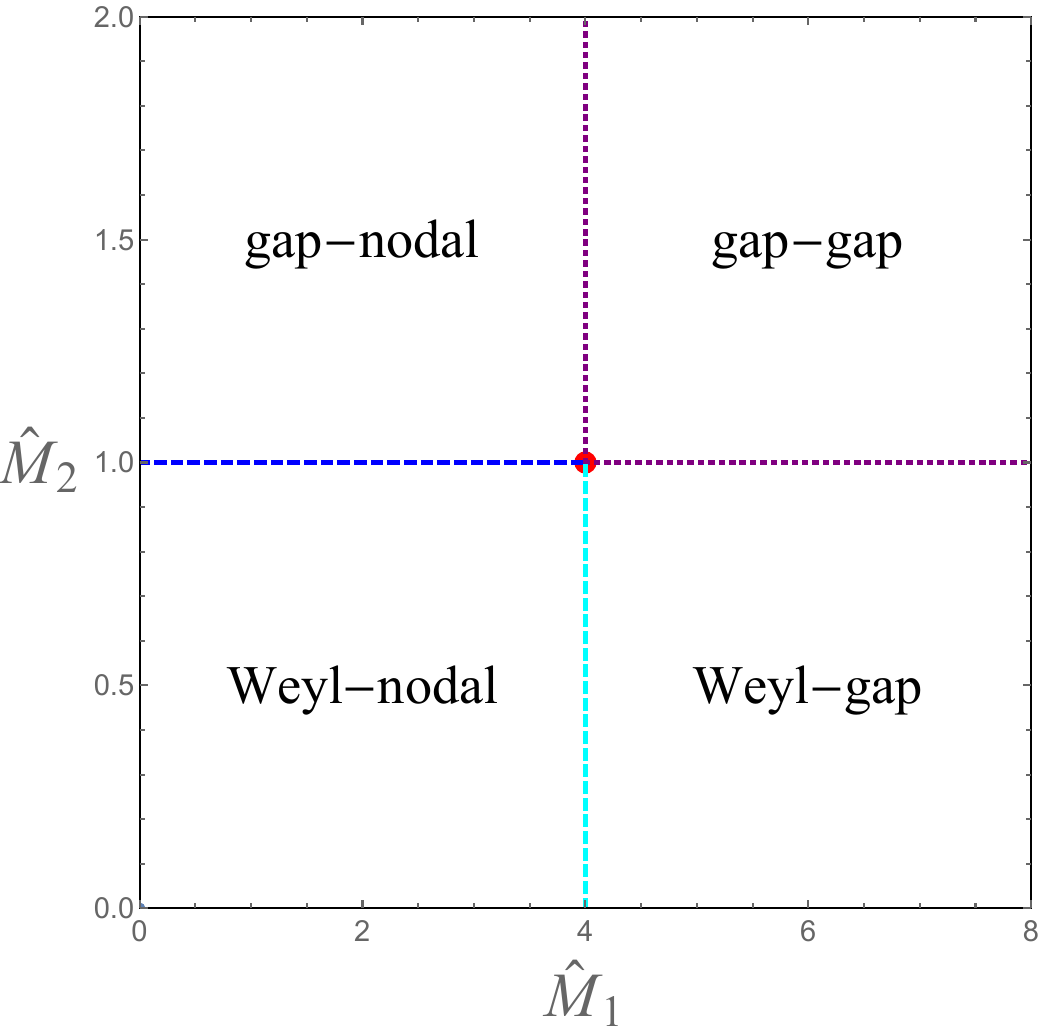}}
            \end{minipage}
            \begin{minipage}{0.49\linewidth}
                \vspace{3pt}
                \centerline{\includegraphics[width=\textwidth,height=\textwidth]{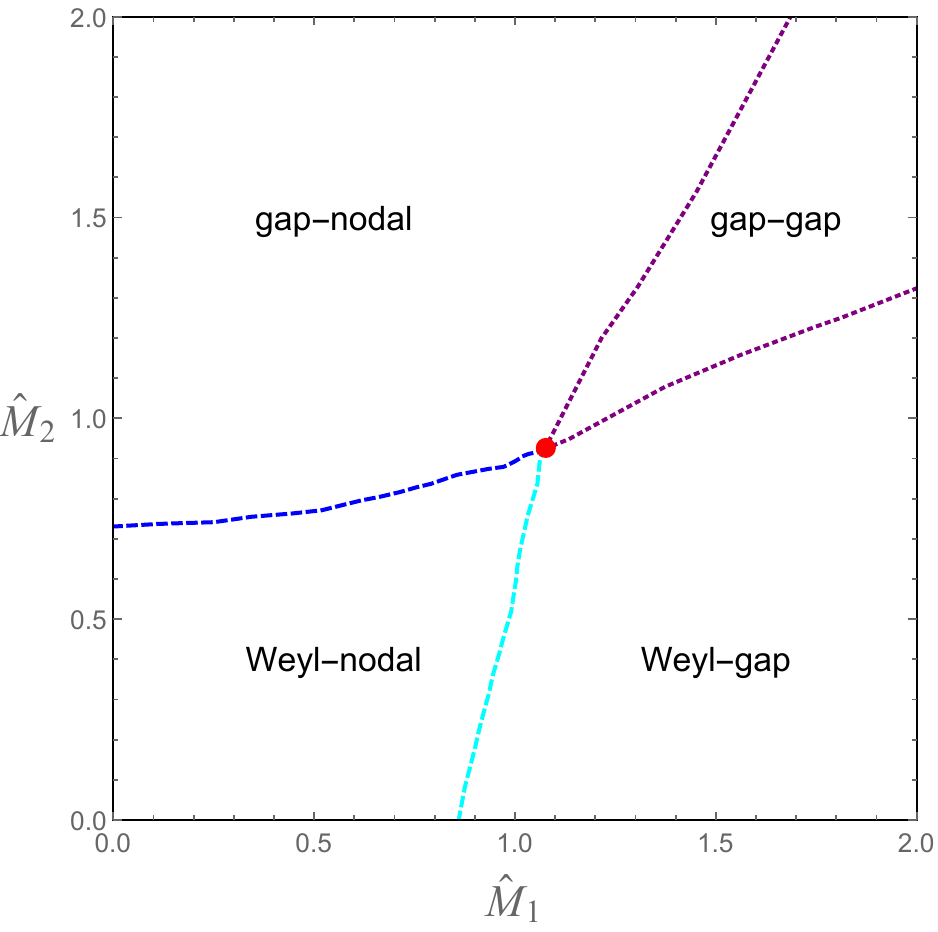}}
            \end{minipage}
            \caption{Left: the phase diagram for the weak coupling Weyl-nodal line  coexisting model \eqref{WeylNodalCoexistLagrangian}. Right: the phase diagram for the holographic Weyl-nodal line coexisting model \eqref{CoexistHolographicModel}. $\hat{M_1}=\frac{M_2}{c}$, $\hat{M_2}=\frac{M_1}{b}$ and $c/b=1$.The red points represent the Critical-Critical phase, where both the Weyl nodes and the nodal line collapse into a single critical point. The cyan dashed line indicates the Weyl-Critical phase, characterized by the vanishing radius of the nodal ring while a pair of Weyl nodes persists. The blue dashed line corresponds to the Critical-Nodal phase, in which the Weyl nodes annihilate into a critical Dirac node, yet the nodal ring remains intact. Finally, the purple dotted lines describe the Critical-Gap (or Gap-Critical) phase, where either the Weyl nodes merge into a critical Dirac point, leaving the nodal line gapped, or the nodal ring collapses to zero radius, while the Weyl nodes become gapped\cite{HolographicWeylNodalChu_2024}.}
            \label{CoexistPhaseDiagram}
        \end{figure}

        The values for ${M_1}/{b},{M_2}/{c}$ are two important dimensionless parameters distinguishing different phases of the system, which play the same role as in the weak coupling coexisting semimetal system. In previous work\cite{HolographicWeylNodalChu_2024}, we have investigated the dependence of the different phases on the parameters $M_1/b, M_2/c$ in the holographic Weyl-nodal line semimetal. The phase diagrams for both the weak coupling model and for the holographic model are shown in Fig. \ref{CoexistPhaseDiagram}. It can be observed that the phase diagram for this strongly coupled coexisting semimetal system is qualitatively the same as that for the weak coupling one.
\section{Topological invariants for coexisting Weyl-nodal line semimetal}
    \label{Section3}
    A non-zero Weyl charge indicates that the Weyl nodes are topologically protected. Similarly, a non-zero $\zeta_1$ ($\zeta_0$)  signifies that the nodal line is topologically protected and remains gapless under small perturbations (protected by the mirror symmetry). Meanwhile, a non-zero $\zeta_2$ ($\widetilde{\zeta}_{2}$) implies that when the nodal line shrinks to a point under perturbations (protected by a mirror symmetry), it will inevitably re-expand into a nodal line. Therefore, without loss of generality, these topological invariants can be chosen to fully characterize the phases of a Weyl-nodal line coexisting semimetal.
    
    In this section, we build upon our previous study of both the weak coupling field theoretical model and holographic constructions for coexisting semimetals, now completing the program by computing their topological invariants(the Weyl charge, $\zeta_0$, $\zeta_1$, $\zeta_2$ and $\widetilde{\zeta}_{2}$). We will first introduce the basic definition and calculation protocols for each topological invariant, carefully specifying their respective integration manifolds (enclosing spheres, linked contours, and nested tori). Then we start with the weak-coupling field theoretical model for the coexisting semimetal, and employ the momentum-space Hamiltonian $H(k)$ derived earlier as in \eqref{WeylNodalCoexistHamiltonian} to visualize the band structure of $H(k)$, obtaining their results of topological invariants.
    
    Then for holographic semimemtals, we first review the bulk probe fermion actions for both Weyl and nodal line holographic models, then synthesize the bulk action for probe fermions in the holographic coexisting Weyl-nodal line semimetal. We then utilize the topological Hamiltonian method to compute the corresponding topological invariants. The topological Hamiltonian emerges naturally from the UV asymptotics behaviors of these probe fermions, allowing us to compute topological invariants parallel to the weak-coupling cases. We calculate the Weyl charge, $\zeta_0$, $\zeta_1$,  $\zeta_2$ and $\widetilde{\zeta}_{2}$ in the Weyl-Nodal phase and plot the corresponding effective band structures. A special behavior, a band crossing ordering interchange phenomena, is also observed in coexisting semimetals, which was previously viewed to be specific for holographic Weyl semimetals.
    
    \subsection{Topological invariants and results in the effective field theoretical model}
        As the coexisting semimetal state has both Weyl nodes and a topological nodal ring, the system possesses topological invariants for both the Weyl nodes and the nodal ring, i.e. the Weyl charge and the $\zeta_0$, $\zeta_1$,  $\zeta_2$ and $\widetilde{\zeta}_{2}$ invariants. In this part we will first introduce the protocols for calculating these topological invariants and then obtain their results in the weak coupling Weyl-nodal line coexisting semimetal model. We can define two sets of topological invariants for the two sectors: the Weyl charge for the Weyl sector and four topological invariants for the nodal line sector, which we will introduce in detail below.
    
        \noindent{\bf The Weyl sector: the Weyl charge.} To begin with, let us discuss how to obtain the Weyl charge in the Weyl-nodal line coexisting semimetal. In previous work\cite{HolographicTopologicalInvariantLiu_2018,HolographicWeylZ2chen2025} we have demonstrated how to get the Weyl charge by integrating the Berry curvature on a closed surface enclosing a Weyl node in the pure Weyl semimetal. The calculation procedure for the Weyl charge in the coexisting semimetal is as follows.
        
\begin{figure}[htbp]
            \begin{minipage}{0.49\linewidth}
                \vspace{3pt}
                \centerline{\includegraphics[width=\textwidth]{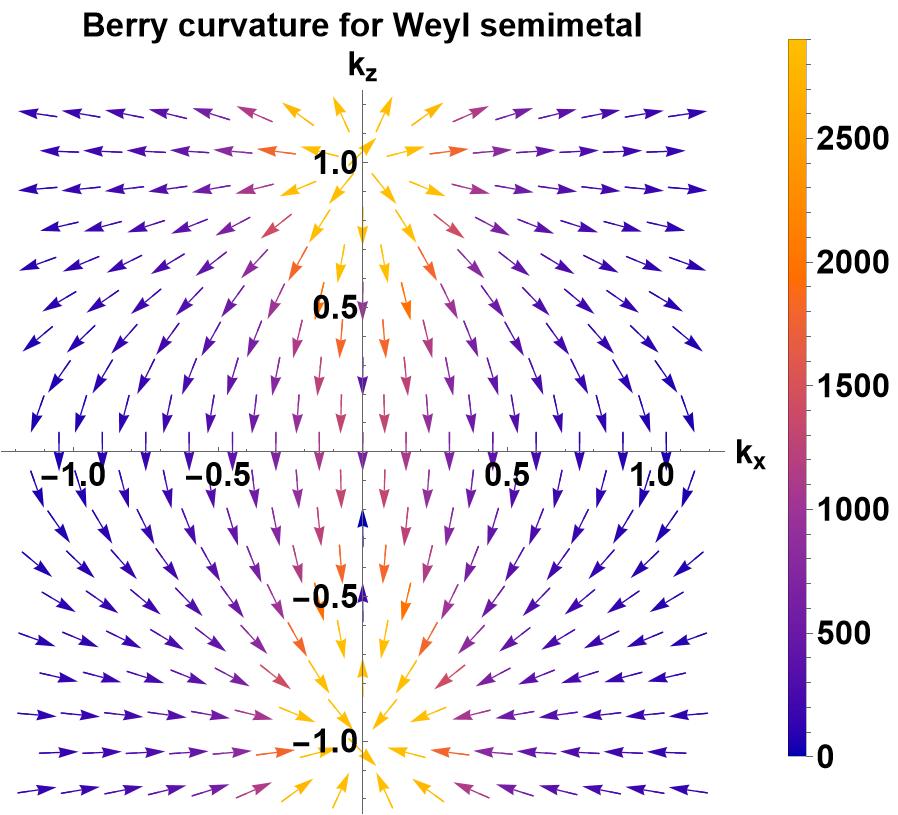}}
            \end{minipage}
            \begin{minipage}{0.49\linewidth}
                \vspace{3pt}
                \centerline{\includegraphics[width=\textwidth]{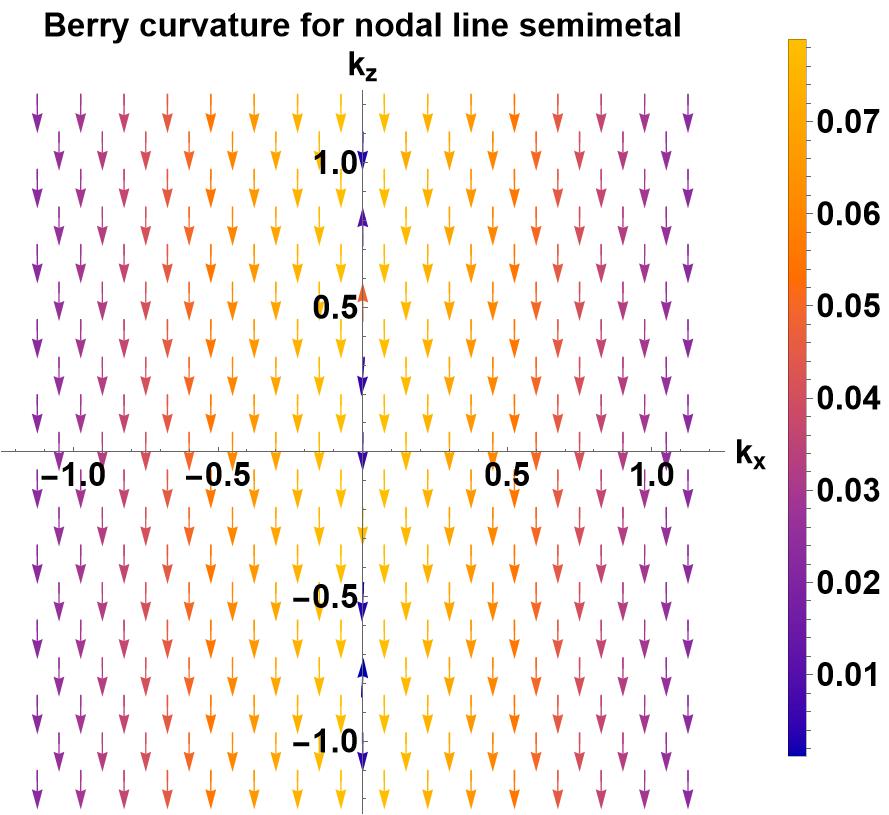}}
            \end{minipage}
            \caption{{The Berry curvature distributions for the pure Weyl semimetal (left panel) and the nodal-line semimetal (right panel) are shown in the $k_y=0$ plane. In the left panel, a distinct source and sink structure is evident, enabling the evaluation of a non-zero Weyl charge through surface integration of the Berry curvature around the source point. In contrast, the right panel exhibits a trivial Berry curvature distribution, confirming that the Weyl charge must vanish in the nodal-line semimetal.}
            }
            \label{PureWeylAndNodalBerryCurvature}
        \end{figure}

        To be specific, we have the Hamiltonian $H(\vec{k})$ for the coexisting semimetal, and consequently a pair of Weyl nodes at $\pm\vec{k}_n=(0,0,\pm k_0)$ in the momentum space. First we need to get the Berry curvature, which will be integrated later. We can solve the eigen-equation $H(\vec{k})\ket{n(\vec{k})}=E_n(\vec{k})\ket{n(\vec{k})}$ to get the eigen-state $\ket{n(\vec{k})}$ whose corresponding eigenvalue $E_n(\vec{k})$ is less than Fermi energy $E_F$, and the Berry curvature $\Omega$ is defined as
        \begin{equation}
            \label{BerryCurvature}
            \Omega=\sum_{E_n\le E_F}id\bra{n(\vec{k})}\wedge d\ket{n(\vec{k})},
        \end{equation}
        where $d$ is exterior differential operator. The corresponding Berry curvature for a Weyl semimetal is shown in the left panel of Fig. \ref{PureWeylAndNodalBerryCurvature}. Then the Weyl charge for the Weyl node at $\vec{k}_n$ can be calculated by integrating the Berry curvature $\Omega$ on a closed surface $\Sigma$ enclosing $\vec{k}_n$ in the momentum space
        \begin{equation}
            \label{WeylCharge}
            \text{Weyl Charge}=\int_\Sigma \sum_{E_n\le E_F}\frac{id\bra{n(\vec{k})}\wedge d\ket{n(\vec{k})}}{2\pi i}.
        \end{equation}
        
        A nonzero Weyl charge reveals that there is a Weyl node enclosed by the integral surface, which is topologically protected and cannot be gapped by small perturbations.
        
        \noindent{\bf The nodal line sector.} For the nodal line sector, depending on whether the nodal ring needs to be protected by a mirror symmetry of $z\to -z$, we could define two sets of topological invariants. In our holographic nodal line system, a mirror symmetry is presented for the background (\ref{NodalHolographicModel}) \footnote{{The holographic model for Weyl-nodal line coexisting semimetal \eqref{CoexistHolographicModel} also preserves the mirror symmetry for the nodal line sector.}}. As topological invariants detect whether the nodal ring is stable under small perturbations, we could require all the perturbations to also obey the mirror symmetry and in this case we  could defined two topological invariants which are mirror symmetry protected topological invariants. In more general cases, if we do not require the perturbations to possess the mirror symmetry, we could define two more general topological invariants, which could be viewed as the generalized version of the two symmetry protected topological invariants. In previous work on holographic calculations of topological invariants for nodal line semimetals \cite{HolographicTopologicalInvariantLiu_2018, HolographicNodalLineSemimetalsLiu_2021}, only one of the four topological invariants has been considered, and in this work, we will calculate all of these four for both the holographic nodal line semimetal and the coexisting semimetal.
        
\begin{figure}[htbp]
            \begin{minipage}{0.48\linewidth}
                \vspace{3pt}
                \centerline{\includegraphics[width=\textwidth,height=0.48\textwidth]{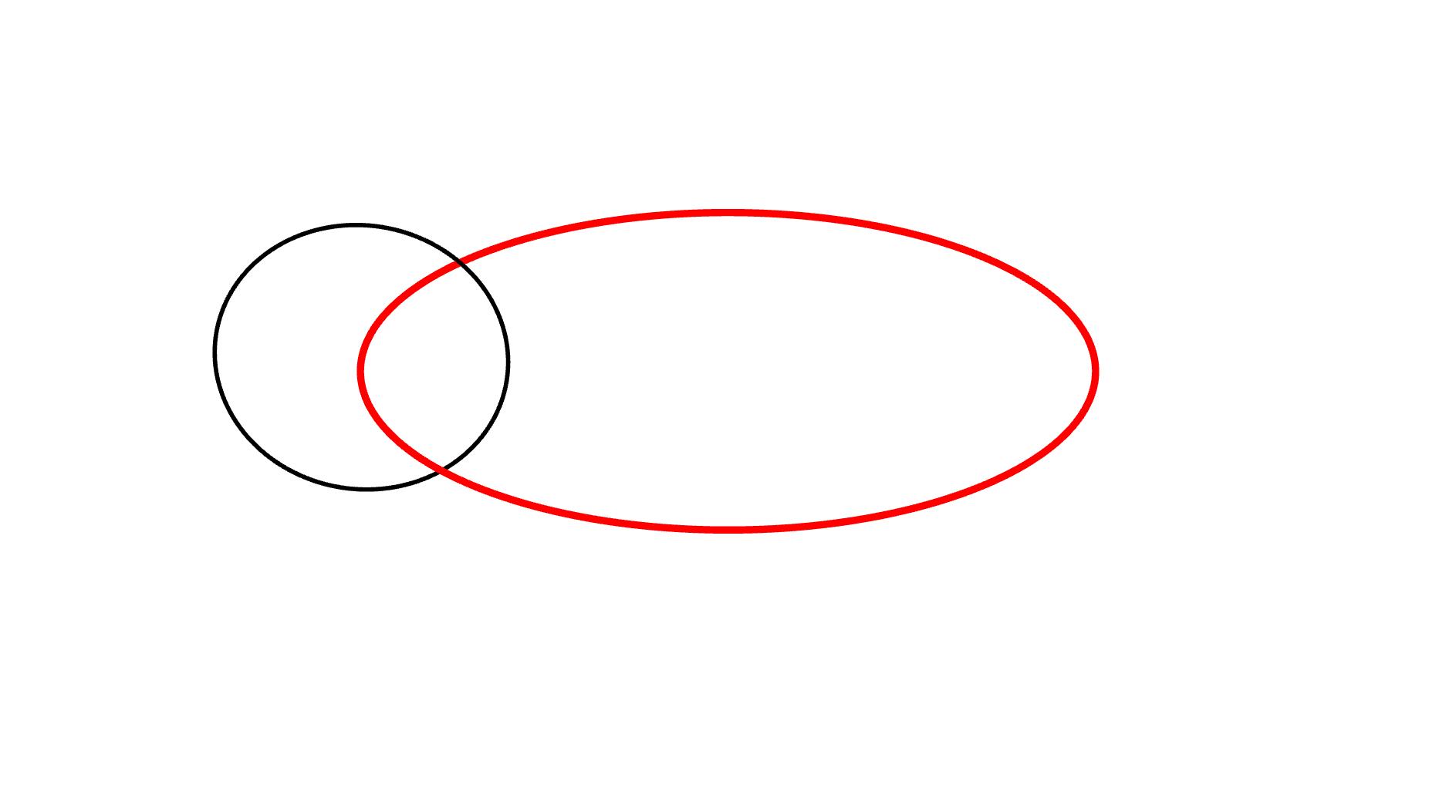}}
            \end{minipage}
            \begin{minipage}{0.48\linewidth}
                \vspace{3pt}
                \centerline{\includegraphics[width=\textwidth,height=0.48\textwidth]{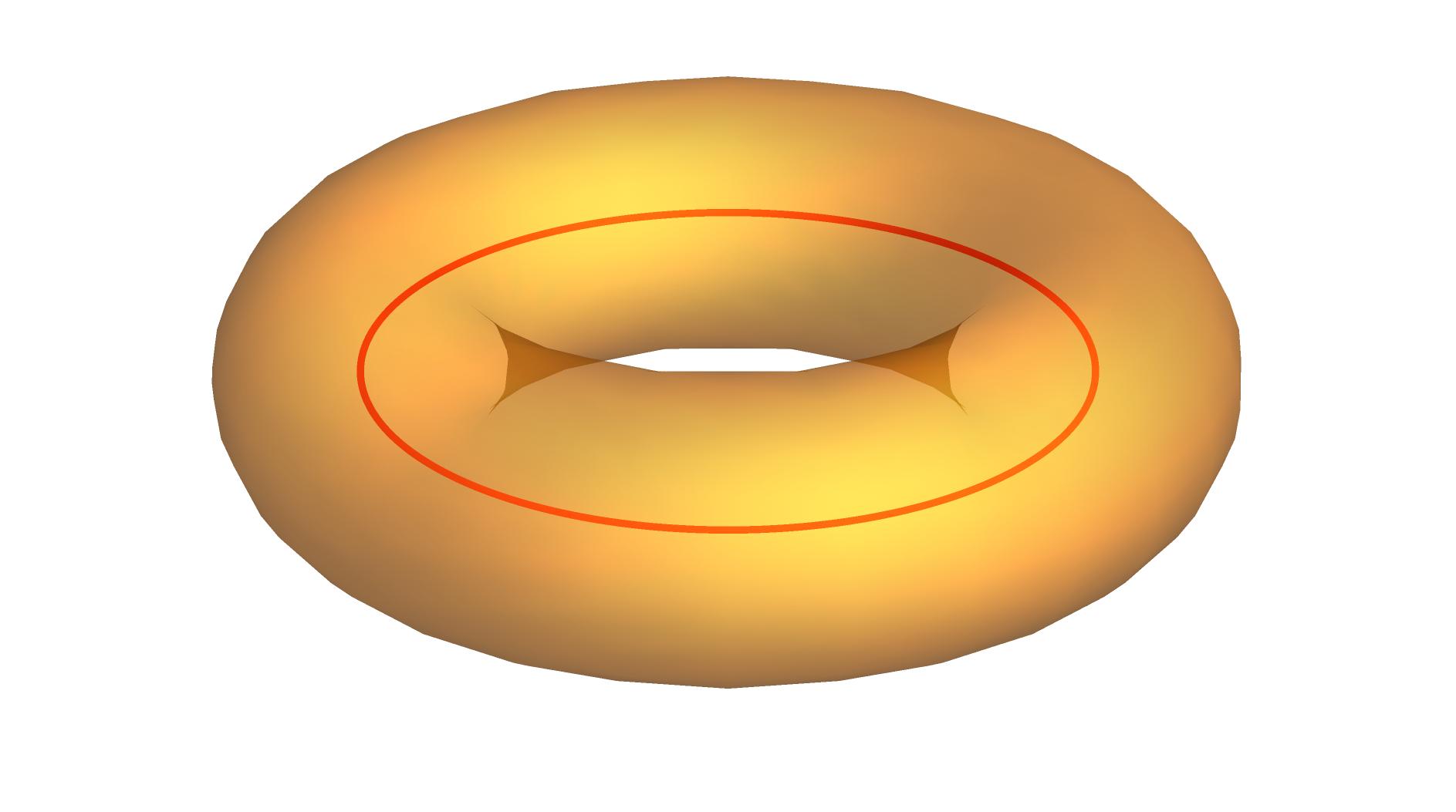}}
            \end{minipage}
            \begin{minipage}{0.48\linewidth}
                \vspace{3pt}
                \centerline{\includegraphics[width=\textwidth,height=0.48\textwidth]{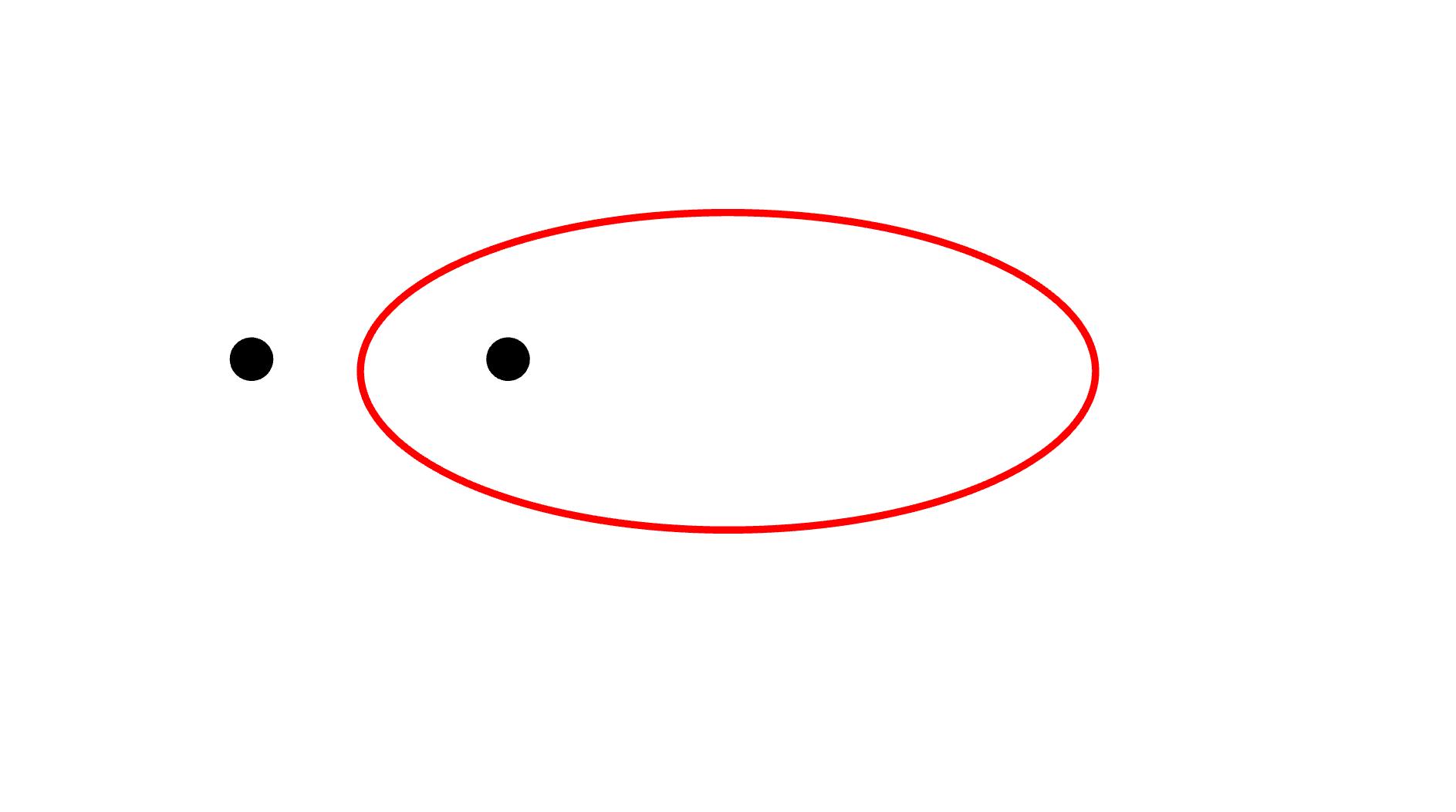}}
            \end{minipage}
            \begin{minipage}{0.48\linewidth}
                \vspace{3pt}
                ~~~~~~~\centerline{\includegraphics[width=\textwidth,height=0.48\textwidth]{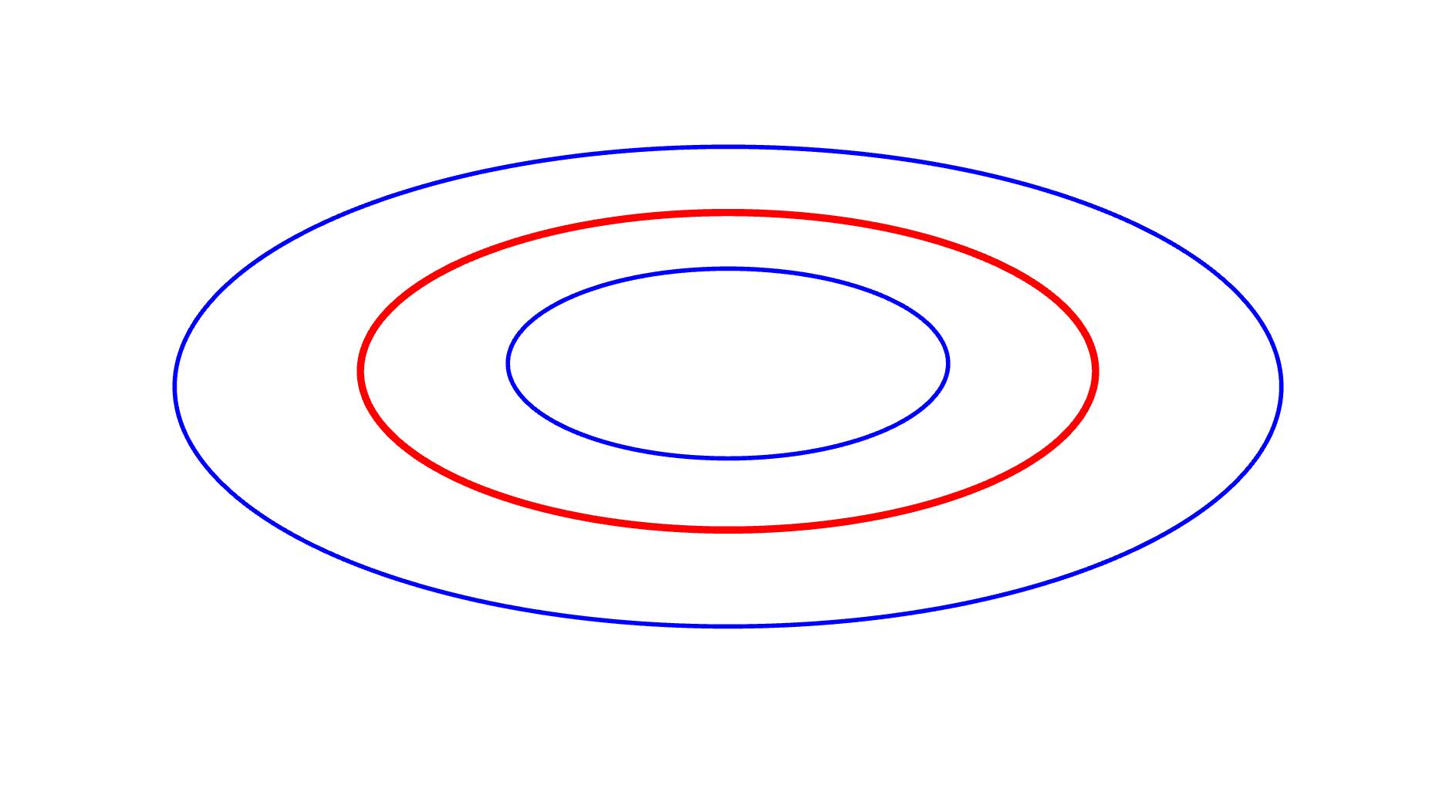}}
            \end{minipage}
            \caption{Manifolds of different dimensions $(\mathbb{S}^0,\mathbb{S}^0\times\mathbb{S}^1,\mathbb{S}^1,\mathbb{T}^2)$ that enclosed the nodal ring (i) a loop $(\mathbb{S}^1)$ linked to the nodal ring to calculate $\zeta_1$. (ii) a torus $(\mathbb{T}^2)$ surrounding the entire nodal ring {to calculate $\zeta_2$}. (iii) two points $(\mathbb{S}^0)$ inside and outside the nodal ring pinned to the mirror plane to calculate $\zeta_0$. (iv) two closed ring $(\mathbb{S}^0\times\mathbb{S}^1)$ inside and out side the nodal ring embedding in the mirror plane to calculate $\widetilde{\zeta}_2$. }
            \label{FigureIntegralManifold}
        \end{figure}

        In the nodal line sector equipped with $TP$ symmetry, the Hamiltonian eigenstates can always be chosen as real-valued wavefunctions, resulting in the trivialization of the conventional Weyl charge defined through Berry curvature integration. In the right panel of Fig. \ref{PureWeylAndNodalBerryCurvature} the Berry curvature of a nodal line semimetal is shown, which is trivial indeed. This fundamental limitation necessitates the introduction of alternative topological invariants, i.e the $\zeta_1$ and $\zeta_2$ invariants, and the $\zeta_0$ and $\widetilde{\zeta}_2$ invariants with additional mirror symmetry as shown in Fig. \ref{FigureIntegralManifold}.
        
\begin{figure}[htbp]
            \begin{minipage}{0.32\linewidth}
                \vspace{3pt}
                \centerline{\includegraphics[width=\textwidth]{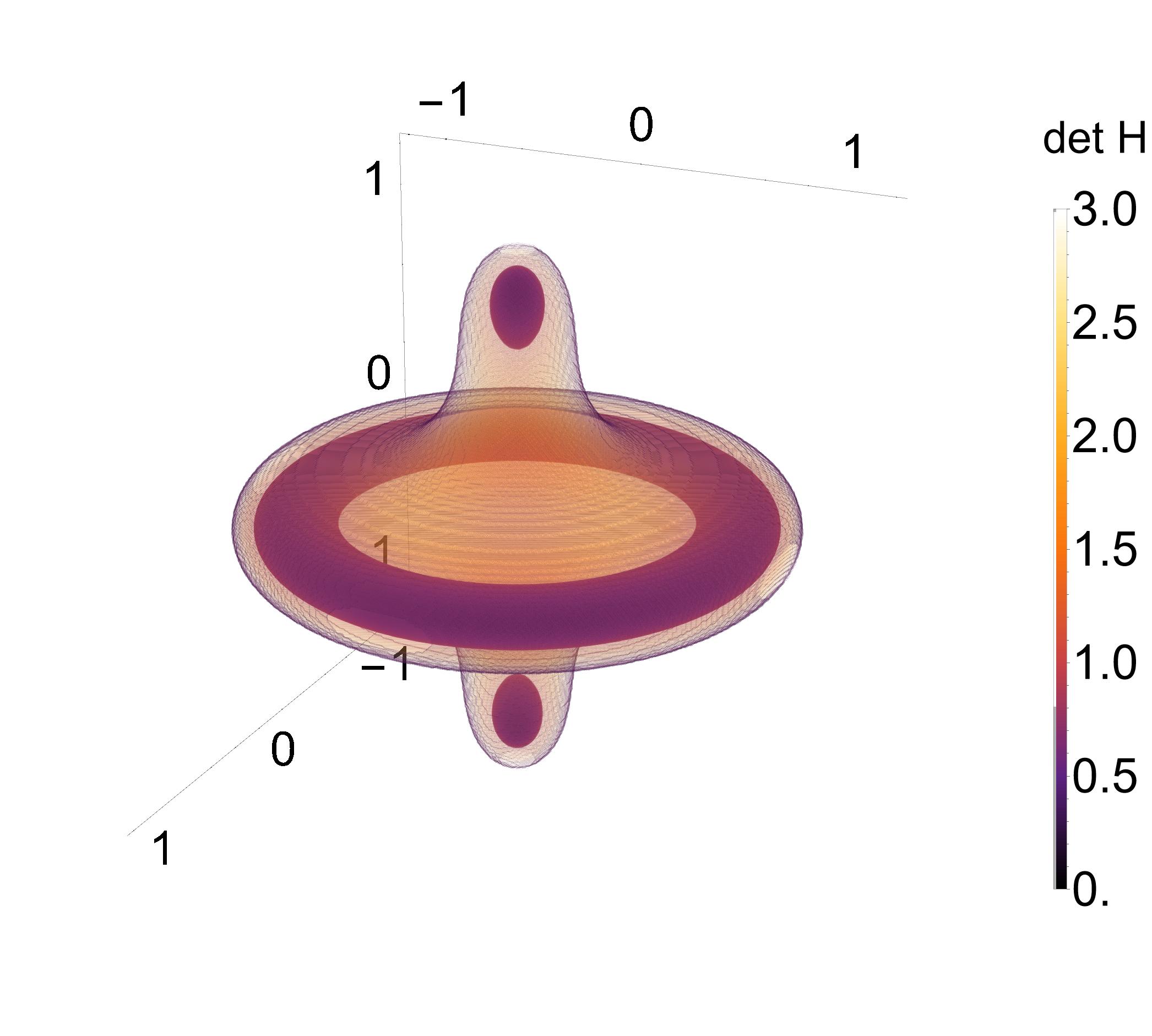}}
            \end{minipage}
            \begin{minipage}{0.32\linewidth}
                \vspace{3pt}
                \centerline{\includegraphics[width=\textwidth]{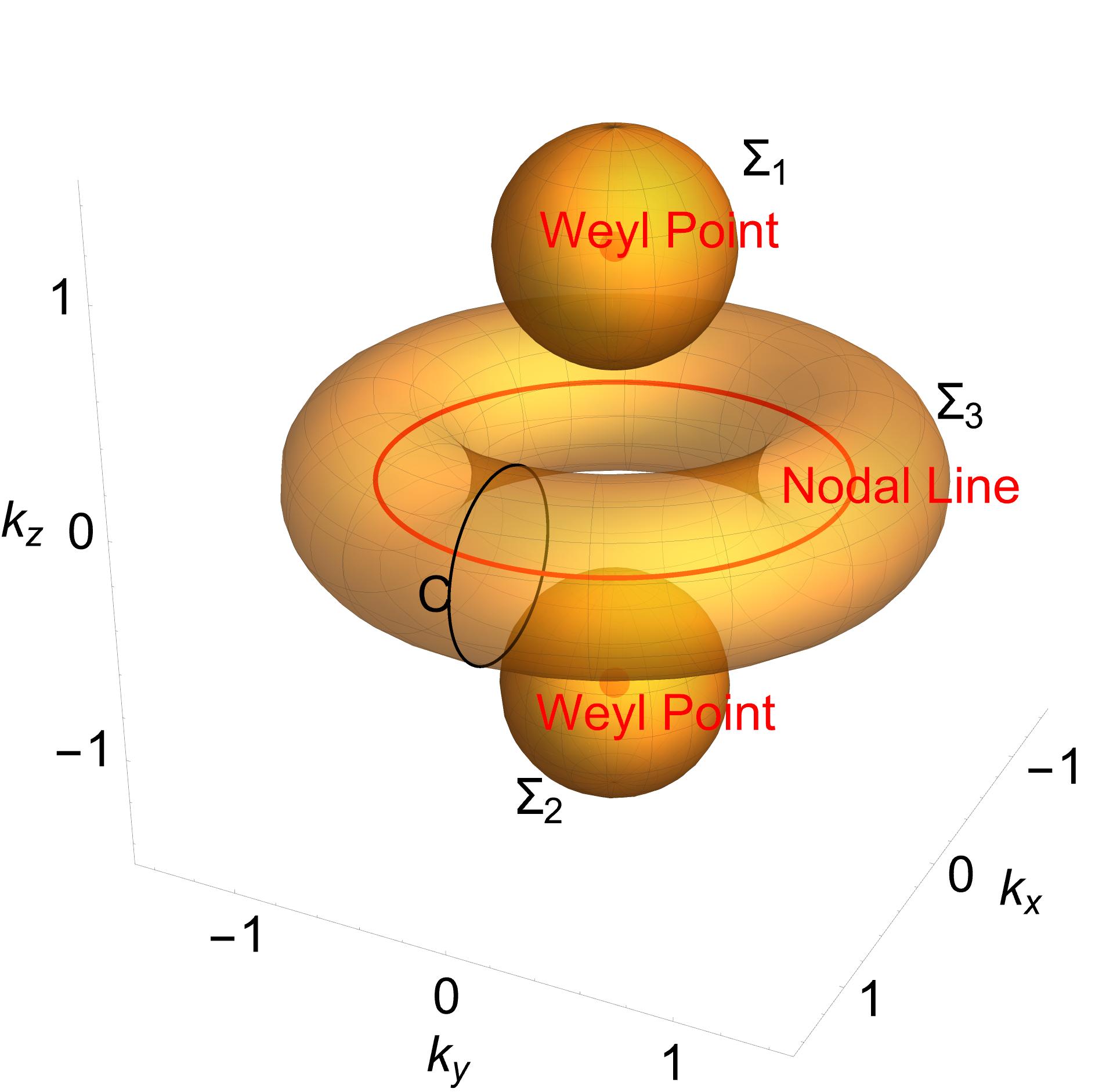}}
            \end{minipage}
            \begin{minipage}{0.32\linewidth}
                \vspace{3pt}
                \centerline{\includegraphics[width=\textwidth]{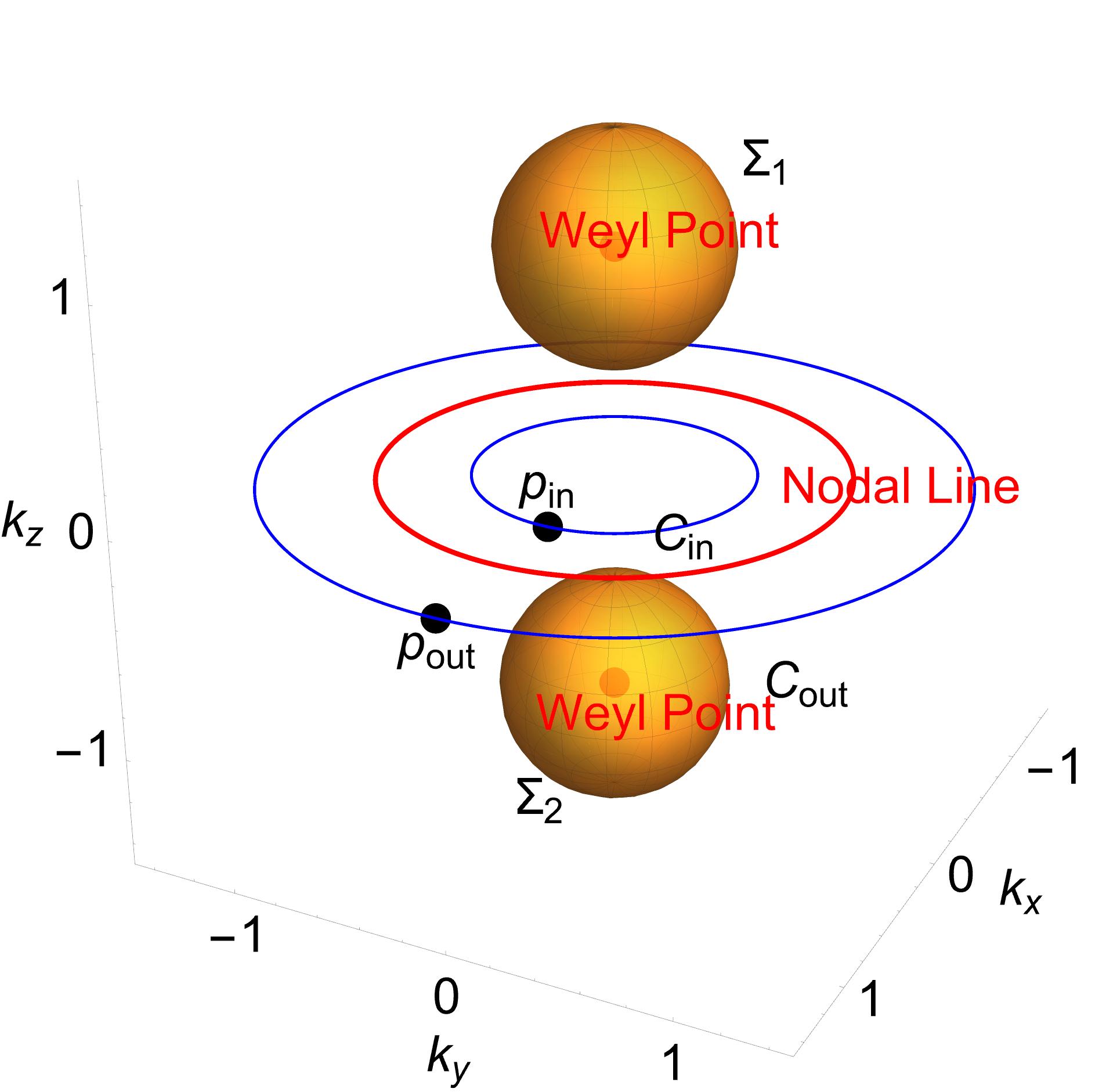}}
            \end{minipage}
            \caption{(i): The density plot of $\det H$ where $H$ is the Hamiltonian for the {weak coupling} Weyl-nodal line coexisting semimetal \eqref{WeylNodalCoexistHamiltonian}: a nodal line in the $k_x-k_y$ plane and a pair of Weyl nodes along the $k_z$ axis. (ii): Integrating the Berry curvature on closed surfaces $\Sigma_1,\Sigma_2$ to get the Weyl charges, which are $\pm 1$ for this weakly coupled field theoretic model. Integrating the Berry connection on the closed curve $C$ to get $\zeta_1$, which is $1\mod 2$, and calculate the winding number for Wannier center on the closed surface $\Sigma_3$ to get $\zeta_2$, which is $0\mod 2$. (iii): Comparing the eigenvalue of mirror operator on the point $p_{in}$ and $p_{out}$ to get $\zeta_0$, which is $1$, {and comparing the average phases for the Wilson loop along the loop $C_{in}$ and the loop $C_{out}$} to get $\widetilde{\zeta}_2$, which is $0 \mod 2$.}
            \label{CoexistWeakCouplingBandStructure}
        \end{figure}

    \begin{itemize}
        \item  Mirror symmetry protected $\zeta_0$ and $\widetilde{\zeta}_2$ invariants. 
        \begin{itemize}
            \item \textbf{The mirror symmetry protected $\zeta_0$ invariant.} Under the protection of mirror symmetry with respect to the $k_x k_y$-plane, the system exhibits enhanced symmetry within the plane, which significantly simplifies the computation of topological invariants while yielding richer topological structures. A prime example is the $\zeta_0$ invariant. The calculation proceeds as follows: (i) select a pair of points $p_1,p_2$ on the mirror invariant plane, situated inside and outside the nodal ring, respectively; (ii) count the number of occupied states $N_1,N_2$ that remain invariant under the mirror reflection operator $M$ at these points; (iii) define $\zeta_0$ as the difference between $N_1$ and $N_2$ $(\zeta_0=N_1-N_2)$. {The integral manifold of $\zeta_0$ is shown in the bottom left panel of Fig. \ref{FigureIntegralManifold}}. If $\zeta_0=0$, the nodal ring is unstable against small perturbations and may open a gap; whereas a non-zero $\zeta_0$ signifies topological protection, ensuring the gap remains closed.
            \item \textbf{The mirror symmetry protected $\widetilde{\zeta}_2$ invariant.} To further investigate the topological properties for nodal line semimetals, we introduce the $\widetilde{\zeta}_2$ topological invariant. {In order to distinguish this $\widetilde{\zeta}_2$ from the general $\zeta_2$ invariant without symmetry protection, here we denote the mirror symmetry protected one using $\widetilde{\zeta}_2$.}
            $\widetilde{\zeta}_2$ is another topological invariant that can provide extra information on topological properties of the nodal ring which $\zeta_0$ alone cannot provide. {We develop a method to calculate this mirror symmetry protected $\widetilde{\zeta}_2$ invariant, inspired by the calculations of similar topologicle invariants in other systems.} The mirror symmetry protected $\widetilde{\zeta}_2$ invariant can be simplified by comparing properties inside and outside the nodal ring on a high-symmetry plane {as shown in the bottom right panel of Fig. \ref{FigureIntegralManifold}}. Specifically, we select two distinct integration loops (circles) on the mirror plane, one inside and one outside the nodal ring, and construct the corresponding non-Abelian Berry connections and Wilson loops. By comparing the average phases for the eigen-values of the Wilson loops on these two paths and considering the constraint from $TP$ symmetry, which dictates that the phase difference must be an integer multiple of $\pi$, we extract the $\widetilde{\zeta}_2$ invariant. 
            \begin{itemize}
                \item  If the phase difference is an even multiple of $\pi$ $(\widetilde{\zeta}_2=0)$, the phase difference is trivial, allowing the nodal ring to be fully contracted to a point and then gapped out through parameter adjustments.
                \item  If the phase difference is an odd multiple of $\pi$ $(\widetilde{\zeta}_2=1)$, it indicates that when system parameters are adjusted to contract the nodal ring to a point, further tuning cannot annihilate it to open a gap; instead, the point will inevitably re-expand into a nodal ring. This robustness stems from the persistent nontrivial topological phase difference between the interior and exterior of the ring, which cannot be smoothly trivialized under continuous deformation.
            \end{itemize}
            
            \quad The procedure for the calculation of the mirror symmetry protected $\widetilde{\zeta}_2$ invariant includes the following three steps.  
            \renewcommand{\labelitemiii}{$\diamond$}
            \begin{itemize}
                \item  (i). Choose two integral loops on the plane where the nodal ring lies, one inside ($C_{in}$ in Fig. \ref{CoexistWeakCouplingBandStructure} b) and one outside ($C_{out}$ in Fig. \ref{CoexistWeakCouplingBandStructure} b) of the nodal ring.

                \item  (ii). Integrate the non-Abelian connection $A$, which is a generalization of the Berry connection, along the two integral loops to get the two corresponding Wilson loops. The non-Abelian Berry connection $A$ with all occupied bands involved is defined by
                    \begin{equation}
                        \label{NonAbelianBerryConnection}
                        A=\begin{pmatrix}
                            \bra{1(\vec{k})}d \ket{1(\vec{k})} & \bra{1(\vec{k})}d \ket{2(\vec{k})} & \cdots &\bra{1(\vec{k})}d \ket{N(\vec{k})}\\
                            \bra{2(\vec{k})}d \ket{1(\vec{k})} & \bra{2(\vec{k})}d \ket{2(\vec{k})} & \cdots &\bra{2(\vec{k})}d \ket{N(\vec{k})}\\
                            \vdots & \vdots & \ddots & \vdots\\
                            \bra{N(\vec{k})}d \ket{1(\vec{k})} & \bra{N(\vec{k})}d \ket{2(\vec{k})} & \cdots &\bra{N(\vec{k})}d \ket{N(\vec{k})}
                        \end{pmatrix},
                    \end{equation}
                    {where $H(\vec{k})\ket{n(\vec{k})}=E_n\ket{n(\vec{k})}$ and $E_1\le E_2\le \cdots E_N \le E_F$ with $E_F$ the Fermi energy.}
                Note that the $\zeta_2$ invariant becomes trivial for a single occupied band, it is essential to examine the non-commutative gauge field contributed by multiple bands.

                \quad For this non-Abelian Berry connection, the calculation of the corresponding Wilson loop should use the path-ordered exponential instead of normal integration. We have
                    \begin{equation}
                        \label{WilsonLoop}
                        W=\mathcal{P}\exp\left(-\int_{C}A\right),
                    \end{equation}
                    where $\mathcal{P}$ is the path-ordered exponential.
                \item (iii). Calculate the eigenvalues of the Wilson loops and obtain the difference of the averaged arguments inside and outside the nodal ring to get the $\widetilde{\zeta}_2$ invariant. Wilson loop $W$ is an orthogonal matrix and all its eigenvalues have the form $e^{i\theta}$, where real-valued $\theta$ is the phase of the eigenvalue of $W(u)$. Denote $\bar{\theta}(u)$ as the average phase for all eigenvalues of $W(u)$, $\widetilde{\zeta}_2$ is defined by
                \begin{equation}
                    \label{NodalZeta2TopologicalInvariant}
                    \widetilde{\zeta}_2=\frac{\Bar{\theta}_{in}-\Bar{\theta}_{out}}{\pi}\mod 2.
                \end{equation}
            \end{itemize}

            {For the nodal line equipped with a non-zero $\widetilde{\zeta}_2$ invariant, there exists a topological protected phase difference between the inside and the outside loop of the nodal ring. When we shrink the nodal ring into a point, we cannot eliminate the phase difference through continuous transition, so the point cannot be gapped and will instead re-expend to a nodal line.}
        \end{itemize}
        \item General topological invariants without symmetry protection: $\zeta_1$ and $\zeta_2$.
        \begin{itemize}
            \item \textbf{The $\zeta_1$ invariant.} The $\zeta_1$ invariant extends the concept of the $\zeta_0$ invariant beyond mirror-symmetric systems, defining a robust topological index protected solely by TP symmetry\footnote{ Equivalently, the $\zeta_0$ invariant could be viewed as the mirror symmetry protected version of  $\zeta_1$.}.   The invariant $\zeta_1$ determines whether the nodal ring is topologically protected, i.e. whether it could be gapped by small perturbations. 
            
            \quad {The precise definition and computational protocol for the $\zeta_1$ invariant are as follows. Consider a topological semimetal hosting a nodal ring. The Berry phase $\zeta_1$ is computed along a momentum space loop $C$ that is topologically linked with the nodal ring, as shown in the upper left panel of  Fig. \ref{FigureIntegralManifold}.}
            \begin{equation}
                \label{NodalZeta1TopologicalInvariant}
                (-1)^{\zeta_1}=\oint_C\omega,~\omega=i\sum_{E_n<E_F}\bra{n(\vec{k})}d\ket{n(\vec{k})},
            \end{equation}
            where $d$ is the exterior differential operator and $\ket{n(\vec{k})}$ is the eigen-state corresponding to the eigenvalue $E_n(\vec{k})$ of the Hamiltonian $H(\vec{k})$ of the nodal line semimetal, i.e. $H(\vec{k})\ket{n(\vec{k})}=E_n(\vec{k})\ket{n(\vec{k})}$. $E_F$ is the Fermi energy and $\omega$ is the Berry connection. Similar to the situation in the Weyl semimetal, a nonzero $\zeta_1$ makes sure that the nodal ring is topologically protected: small perturbations cannot gap the nodal ring directly.
            \item \textbf{The $\zeta_2$ invariant.} In systems protected solely by $TP$ symmetry without mirror symmetry, the $\zeta_2$ invariant remains essential for characterizing the topology of the nodal-line structure. 
            However, within this symmetry setting, $\zeta_2$ cannot be obtained by simply comparing the average phase differences at high-symmetry points; instead, it must be computed by evaluating the winding number of the Wannier bands. Despite the different computational approach, its topological significance parallels the mirror-symmetric case: a non-zero $\zeta_2$ implies the impossibility of contracting the nodal ring into a point and then gapping it out through continuous parameter tuning, whereas a vanishing $\zeta_2$ indicates that this process is topologically allowed.
            
            \quad  Specifically, one must construct a sphere or a torus in the momentum space that encloses the nodal ring as shown in the upper right panel of Fig. \ref{FigureIntegralManifold}, and then compute the evolution of the Wannier centers on this torus. For a torus surface in the momentum space, it can always be parameterized as follows
            \begin{equation}
                \begin{aligned}
                    k_x(u,v)&=(R+r\cos v)\cos u,\\
                    k_y(u,v)&=(R+r\cos v)\sin u,\\
                    k_z(u,v)&=R\sin v.
                \end{aligned}
            \end{equation}
        
            \quad For a fixed parameter $v$, an integration contour is chosen with $u$ varying from $0$ to $2\pi$. The corresponding Wilson loop $W(v)$ is computed using \eqref{NonAbelianBerryConnection} and \eqref{WilsonLoop}. Since $W(v)$ is a unitary matrix, all its eigenvalues must take the form $e^{2\pi i \theta(v)}$, where $\theta(v)$ is a real phase function. The $\zeta_2$ invariant is then obtained by calculating the winding number of this phase function \cite{NonAbelianConnectionYu_2011}
           \begin{equation}
                \zeta_2=\left[\theta(0)+\int_0^{2\pi}\frac{d\theta(v)}{dv}dv-\theta(2\pi)\right]\mod 2\pi.
            \end{equation}
        
            \quad The $\zeta_2$ invariant defined here shares the same physical interpretation as its counterpart in mirror-symmetric systems: a non-zero $\zeta_2$ signifies that the nodal ring, upon being contracted to a point, cannot be gapped out due to topological protection; whereas a vanishing $\zeta_2$ indicates that the nodal ring can be topologically trivialized through contraction followed by gap opening.
        \end{itemize}
        
    \end{itemize}

    \begin{figure}[htbp]
            \centering
            \begin{minipage}[t]{\linewidth}
                \centerline{\includegraphics[width=\textwidth]{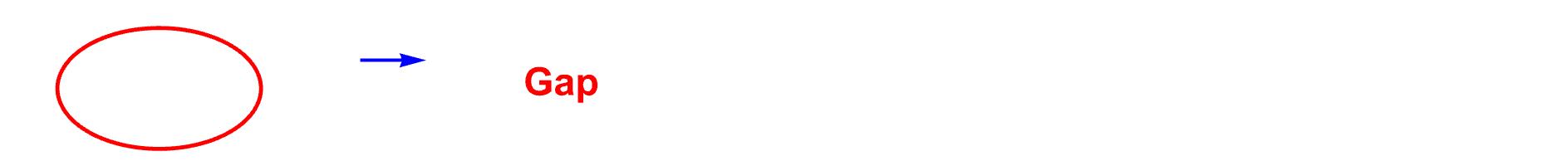}}
            \end{minipage}
            \begin{minipage}[t]{\linewidth}
                \centerline{\includegraphics[width=\textwidth]{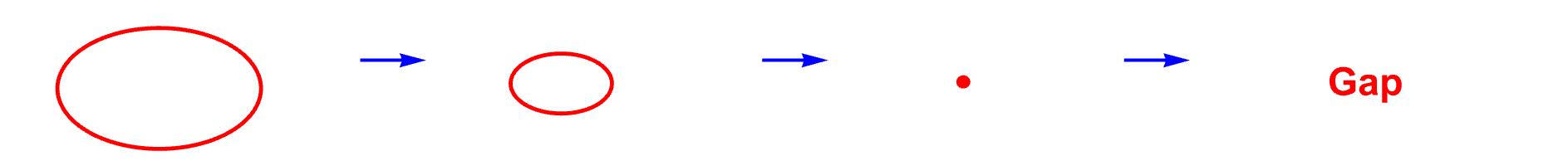}}
            \end{minipage}
            \begin{minipage}[t]{\linewidth}
                \centerline{\includegraphics[width=\textwidth]{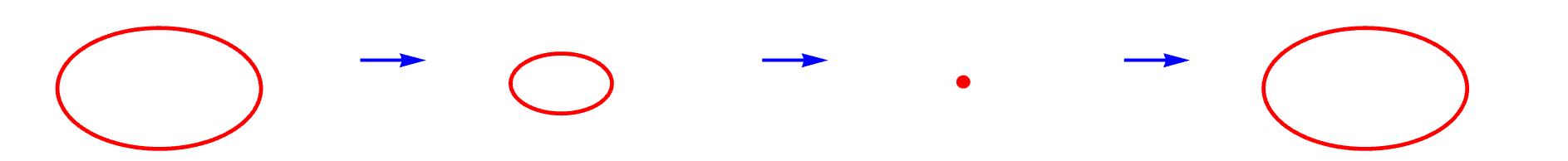}}
            \end{minipage}
            \caption{(i)When $\zeta_1=0$ the nodal ring is not stable and will be gapped under small perturbations, in this situation the $\zeta_2$ invariant must be $0$. (ii) When $\zeta_1=1, \zeta_2=0$ the nodal ring can be gapped by modifying the parameters to shrink the nodal ring to a critical point. (iii) When $\zeta_1=1, \zeta_2=1$ the nodal ring will not be gapped by shirnking the nodal ring to a critical point but re-expand to a nodal ring.}
            \label{FigureTopoligicalInvariantMeaning}
        \end{figure}
    
        {We summarize all the physical roles of these four topological invariants in Fig. \ref{FigureTopoligicalInvariantMeaning}, where $\zeta_0$ and $\widetilde{\zeta}_2$ play similar roles as $\zeta_1$ and $\zeta_2$ when having a protecting mirror symmetry. Note that in general, if perturbations that obey the required symmetry could destroy the topological state, resulting in a trivial symmetry protected topological invariant, this would indicate that more general perturbations that do not obey the required symmetry would definitely destroy the topology, too. Thus, when the symmetry protected $\widetilde{\zeta}_2$ is zero, this indicates that the general topological invariant $\zeta_2$ should also be zero, confirming our numerical results.}   
      
        As we have introduced the physical significance and computational protocol of all five topological invariants (Weyl charge/$\zeta_0$/$\zeta_1$/$\zeta_2$/$\widetilde{\zeta}_{2}$) in Weyl-Nodal line coexisting semimetals, we now first apply the computation procedures to weak coupling coexisting semimetals here and then subsequently extend the calculations to strongly correlated holographic scenarios in the next subsection.

              The Weyl nodes and the nodal ring in the weak coupling Weyl-Nodal line coexisting semimetal are shown in the density plot of $\det H$ in Fig. \ref{CoexistWeakCouplingBandStructure}.
        To determine the topological invariants in the weak coupling Weyl-Nodal line coexisting semimetal, we adopt the integration paths/surfaces as illustrated in  Fig. \ref{CoexistWeakCouplingBandStructure}. 
        By applying this computational protocol to the weak coupling Weyl-nodal line coexisting Hamiltonian \eqref{WeylNodalCoexistHamiltonian}, we obtain the following results for topological charges:  (1) the Weyl charges for the two nodes along the $k_z$-axis are  $\pm 1$; (2) the $\zeta_1$ invariant for the nodal line equals 1; (3) the $\zeta_2$ invariant for the nodal line vanishes; (4) the mirror symmetry protected $\zeta_0$ is $1$; (5) the mirror symmetry protected $\widetilde{\zeta}_2$ is $0$. These calculations unambiguously demonstrate that \eqref{WeylNodalCoexistHamiltonian} indeed describes a genuine Weyl-nodal line coexisting semimetal. This indicates that the weakly coupled Weyl-nodal line coexisting semimetal possesses both topologically protected Weyl nodes and a nodal line. These features remain robust against {both mirror symmetry protected and general} small perturbations, maintaining their gapless property. Furthermore, under continuous parameter variation, the nodal line undergoes a topological transition: it first shrinks to a critical point before completely vanishing.

    \subsection{Fermionic spectrum and topological invariants for holographic coexisting semimetals}
        In this section, we will introduce the topological Hamiltonian method to obtain the topological invariants (Weyl charge/$\zeta_0$/$\zeta_1$/$\zeta_2$/$\widetilde{\zeta}_{2}$)  for the holographic Weyl-nodal line coexisting semimetal. We will first review the bulk action of probe fermions in the holographic Weyl semimetal and the holographic nodal line semimetal. Then we will develop the bulk action of probe fermions in the holographic coexisting semimetal. We will analyze the UV asymptotic behavior of probe fermions, using which we can construct the Green's function for the holographic coexisting semimetal. Finally, from the zero frequency Green's function we can get the topological Hamiltonian, using which we can get the topological invariants.
        
        \noindent{\bf Probe fermions on the holographic Weyl semimetal.}
        We employ a pair of four-component probe fermions in the bulk to calculate the dual Green's function and the corresponding topological Hamiltonian for the holographic Weyl semimetal. 
        The bulk action for probe fermions in the holographic Weyl semimetal is\cite{HolographicWeylSemimetalsPlantz_2018}
        \begin{equation}
            \label{WeylProbeFermionBulkAction}
            \begin{aligned}
                S_{{f-W 1}}&=i\int d^5x\sqrt{-\det g}\bar{\Psi}_{W1}(\Gamma^a D_a-iqA_a\Gamma^a-m)\Psi_{W1},\\
                S_{{f-W 2}}&=i\int d^5x\sqrt{-\det g}\bar{\Psi}_{W1}(-\eta\Phi\mathbb{I}_4)\Psi_{W2},\\
                S_{{f-W 3}}&=i\int d^5x\sqrt{-\det g}\bar{\Psi}_{W2}(\Gamma^a D_a+iqA_a\Gamma^a+m)\Psi_{W2},\\
                S_{{f-W 4}}&=i\int d^5x\sqrt{-\det g}\bar{\Psi}_{W2}(-\eta^*{\Phi}^*\mathbb{I}_4)\Psi_{W1},\\
                S_{f-W}&=S_{{f-W1}}+S_{{f-W2}}+S_{{f-W3}}+S_{{f-W4}},
            \end{aligned}
        \end{equation}
        where $\mathbb{I}_4$ is $4\times 4$ identity matrix , $q$ is the coupling constant for axial gauge field $A$, which has been introduced in \eqref{WeylHolographicModel}, $\eta$ is the coupling constant for Yukawa term, while $\phi$ is the same in \eqref{WeylHolographicModel}, and $\Psi_{W1},\Psi_{W2}$ are two four-component spinors with $D_a$ the corresponding covariant derivative which has been defined in Appendix \ref{AppendixSpinConnection}. $m$ is the mass for probe fermions. {In the holographic duality between five-dimensional gravity and four-dimensional boundary field theory, the dimensional mismatch between bulk and boundary spinor representations dictates that Dirac fermions in the bulk always correspond to massless chiral fermions on the boundary. To construct massive fermionic excitations and derive non-trivial fermion spectral functions in the boundary theory, we introduce a pair of boundary fermion fields $\Psi_1,\Psi_2$ coupled to an auxiliary scalar field $\Phi$ via Yukawa interaction $\Phi\bar{\Psi}_1\Psi_2+h.c.$ thereby generating a  massive Dirac fermion on the boundary.}
       
       \noindent{\bf Probe fermions on the holographic nodal-line semimetal.} The bulk action for probe fermions in nodal line semimetals is\cite{HolographicNodalLineSemimetalsLiu_2021}
        \begin{equation}
            \label{NodalProbeFermionBulkAction}
            \begin{aligned}
                S_{f-n1}&=i\int d^5x\sqrt{-\det g}\bar{\Psi}_{n1}(\Gamma^aD_a-m)\Psi_{n1},\\
                S_{f-n2}&=i\int d^5x\sqrt{-\det g}\bar{\Psi}_{n1}(-\eta_1\Phi_2\mathbb{I}_4+\eta_2B_{ab}\Gamma^{ab}\gamma^5)\Psi_{n2},\\
                S_{f-n3}&=i\int d^5x\sqrt{-\det g}\bar{\Psi}_{n2}(\Gamma^aD_a+m)\Psi_{n2},\\
                S_{f-n4}&=i\int d^5x\sqrt{-\det g}\bar{\Psi}_{n2}(-{\eta_1}^*{\Phi_2}^*\mathbb{I}_4-{\eta_2}^*{B_{ab}}^*\Gamma^{ab}\gamma^5)\Psi_{n1},\\
                S_{f-n}&=S_{f-n1}+S_{f-n2}+S_{f-n3}+S_{f-n4},
            \end{aligned}
        \end{equation}
        where $\mathbb{I}_4$ is $4\times 4$ identity matrix and $\Psi_{n1},\Psi_{n2}$ are two four-component spinors with $D_a$ the corresponding covariant derivative. $\eta_1$, $\eta_2$ are the coupling constants for the Yukawa term and the 2-form field $B$. The 2-form field $B$ and the scalar field $\Phi$ have been defined in \eqref{NodalHolographicModel}. {Similar to holographic Weyl semimetals, the employment of a pair of probe fermions which couple to a scalar field is required in this framework to obtain a massive boundary fermion.}
        
        \noindent{\bf Probe fermions on the holographic coexisting semimetal.}
        After rewiewing holographic Weyl semimetals and nodal line semimetals, we can continue to discuss the holographic Weyl-nodal line coexisting semimetal. Unlike holographic Weyl or nodal line semimetal, the coexistence phase requires a pair of eight component spinors to construct the topological Hamiltonian, reflecting the partial breaking of $TP$ symmetry. This aligns precisely with weak-coupling field theory models, where pure Weyl/nodal line systems are described by 4-component fermions and coexisting semimetal necessitate eight component fermions.

        Hinted from the Lagrangian for the weak coupling coexisting topological semimetal \eqref{WeylNodalCoexistLagrangian}, the bulk action for the two eight-component probe fermions $\Psi_1,\Psi_2$ in the holographic Weyl-nodal line coexisting semimetal can be written as
        \begin{equation}
            \label{CoexistProbeFermionBulkAction}
            \begin{aligned}
                S_1&=i\int d^5x\sqrt{-\det g}\bar{\Psi}_1(\Gamma^a D_a-iqA_a\Gamma^a-m)\oplus(\Gamma^aD_a-m)\Psi_1,\\
                S_2&=i\int d^5x\sqrt{-\det g}\bar{\Psi}_1(-\eta\Phi_1\mathbb{I}_4)\oplus(-\eta_1\Phi_2\mathbb{I}_4+\eta_2B_{ab}\Gamma^{ab}\gamma^5)\Psi_2,\\
                S_3&=i\int d^5x\sqrt{-\det g}\bar{\Psi}_2(\Gamma^a D_a+iqA_a\Gamma^a+m)\oplus(\Gamma^aD_a+m)\Psi_2,\\
                S_4&=i\int d^5x\sqrt{-\det g}\bar{\Psi}_2(-\eta^*{\Phi_1}^*\mathbb{I}_4)\oplus(-{\eta_1}^*{\Phi_2}^*\mathbb{I}_4-{\eta_2}^*{B_{ab}}^*\Gamma^{ab}\gamma^5)\Psi_1,\\
                S&=S_1+S_2+S_3+S_4,
            \end{aligned}
        \end{equation}
        where $\mathbb{I}_4$ is the $4\times 4$ identity matrix, $q,\eta,\eta_1,\eta_2$ are all coupling constants which can be chosen to be real, and $\Gamma^a, D_a$ is defined in appendix \ref{AppendixSpinConnection}, $\Gamma^{ab}=\frac{i}{2}[\Gamma^a,\Gamma^b ]$. {Similar to the holographic Weyl semimetals and holographic nodal line semimetals, we adopt a pair of probe fermions $\Psi_1,\Psi_2$ to produce a boundary massive Dirac fermion. To guarantee the topological coexistence of Weyl points and nodal lines, the introduced probe fermions must be eight-component spinors (rather than four-component ones). This extended spinor representation provides the necessary degrees of freedom to simultaneously accommodate both types of topological defects in the system.} To analyze the behavior of probe fermions, we compute their equations of motion from the bulk action. After performing Fourier transformation on the bulk fermion field $\Psi_l=\psi_l(r) e^{-i\omega t+i k_x x+ik_y y +ik_z z}$, the equations for the fermions can be written explicitly
        \begin{equation}
            \label{CoexistProbeFermionEquationOfMotion}
            \begin{aligned}
                \left(\sqrt{u}\Gamma^r\partial_r+ik_\mu\Gamma^\mu-iq\frac{A_z}{\sqrt{h}}\Gamma^z-m\right)\oplus\left(\sqrt{u}\Gamma^r\partial_r+ik_\mu\Gamma^\mu-m\right)&\psi_1\\
                +\left(-\eta\phi_1\mathbb{I}_4\right)\oplus\left(-\eta_1\phi_2\mathbb{I}_4+2\eta_2\left(\frac{B_{xy}}{f}\Gamma^{xy}+i\frac{B_{tz}}{\sqrt{uh}}\Gamma^{tz}\right)\gamma^5\right)&\psi_2=0,\\
                \left(\sqrt{u}\Gamma^r\partial_r+ik_\mu\Gamma^\mu+iq\frac{A_z}{\sqrt{h}}\Gamma^z+m\right)\oplus\left(\sqrt{u}\Gamma^r\partial_r+ik_\mu\Gamma^\mu+m\right)&\psi_2\\
                +\left(-\eta\phi_1\mathbb{I}_4\right)\oplus\left(-\eta_1\phi_2\mathbb{I}_4-2\eta_2\left(\frac{B_{xy}}{f}\Gamma^{xy}-i\frac{B_{tz}}{\sqrt{uh}}\Gamma^{tz}\right)\gamma^5\right)&\psi_1=0,
            \end{aligned}
        \end{equation}
        where $k_\mu=(-\frac{\omega}{\sqrt{u}},\frac{k_x}{\sqrt{f}},\frac{k_y}{\sqrt{f}},\frac{k_z}{\sqrt{h}})$. The construction of the topological Hamiltonian relies on the UV behavior of the probe fermions. To this end, we observe that the equations of motion exhibit the following asymptotic form as $r\to \infty$. According to the UV behavior of the background geometry \eqref{BackGroundBoundaryBehaviour}, we can get the leading order of the equations \eqref{CoexistProbeFermionEquationOfMotion} near the UV boundary
        \begin{equation}
            \label{CoexistProbeFermionBoundaryEquation}
            (r\Gamma^r\oplus\Gamma^r\partial_r-m)\psi_1=0,~
            (r\Gamma^r\oplus\Gamma^r\partial_r+m)\psi_2=0.
        \end{equation}
        
        Without loss of generality, we take the chiral-Weyl representation for Gamma matrices, which can be found in appendix \ref{AppendixSpinConnection}, the solution for \eqref{CoexistProbeFermionBoundaryEquation} can be written explicitly
        \begin{equation}
            \label{CoexistProbeFermionBoundaryBehaviour}
            \begin{aligned}
                \psi_1&=\begin{pmatrix}a_1 r^{m}, & a_2 r^{m}, & a_3 r^{-m}, & a_4 r^{-m}, & a_5 r^{m}, & a_6 r^{m}, & a_7 r^{-m}, & a_8 r^{-m}\end{pmatrix}^\mathbf{T},\\
                \psi_2&=\begin{pmatrix}b_1 r^{-m}, & b_2 r^{-m}, & b_3 r^{m}, & b_4 r^{m}, & b_5 r^{-m}, & b_6 r^{-m}, & b_7 r^{m}, & b_8 r^{m}\end{pmatrix}^\mathbf{T}.
            \end{aligned}
        \end{equation}
        
        According to the holographic dictionary, we construct the topological Hamiltonian by identifying (1) Source terms: coefficients of the $r^m$ terms in the UV boundary expansion of probe fermions (2) Response terms: coefficients of the $r^{-m}$ terms.
        For convenience, we denote the UV behavior in \eqref{CoexistProbeFermionBoundaryBehaviour} as $\psi_1=a\cdot \begin{pmatrix}r^{m}, & r^{m}, & r^{-m}, & r^{-m}, & r^{m}, & r^{m}, & r^{-m}, & r^{-m}\end{pmatrix}^\mathbf{T}$ and $\psi_2=b\cdot \begin{pmatrix}r^{-m}, & r^{-m}, & r^{m}, & r^{m}, & r^{-m}, & r^{-m}, & r^{m}, & r^{m}\end{pmatrix}^\mathbf{T}$ for simplicity, where $\cdot$ denotes element-wise product and $a=\begin{pmatrix}a_1, & a_2, & a_3, & a_4, & a_5, & a_6, & a_7, & a_8\end{pmatrix}^\mathbf{T}, b=\begin{pmatrix}b_1, & b_2, & b_3, & b_4, & b_5, & b_6, & b_7, & b_8\end{pmatrix}^\mathbf{T}$ are determined by IR boundary conditions. 
        
        For {any set of source $\psi_{source}$ and response $\psi_{response}$ under a given IR boundary condition, they can be expressed by $a,b$,} $\psi_{source}=(1+\Gamma^r\oplus\Gamma^r)a/2+(1-\Gamma^r\oplus\Gamma^r)b/2$, $\psi_{response}=-(1+\Gamma^r\oplus\Gamma^r)b/2+(1-\Gamma^r\oplus\Gamma^r)a/2$, and satisfy the equation $i\Gamma^0 \psi_{response}=G\psi_{source}$, where $G$ is the Green's function, so we have to get eight sets of linearly independent sources and responses to compute the Green's function $G$. To solve \eqref{CoexistProbeFermionEquationOfMotion} numerically we need to first fix $\omega$ and $k=(k_x,k_y,k_z)$, and then by solving \eqref{CoexistProbeFermionEquationOfMotion} with eight linearly independent IR boundary conditions, eight sets of $a^i,b^i$ can be obtained. The source matrix, response matrix and Green's function can be obtained
        \begin{equation}
            \label{CoexistProbeFermionGreenFunction}
            \begin{aligned}
                M_{source}&=\frac{1+\Gamma^r\oplus\Gamma^r}{2}\begin{pmatrix}a^{\mathbf{I}}, & a^{\mathbf{II}}, & a^{\mathbf{III}}, & a^{\mathbf{IV}}, & a^{\mathbf{V}}, &a^{\mathbf{VI}}, & a^{\mathbf{VII}}, & a^{\mathbf{VIII}}\end{pmatrix}\\
                &+\frac{1-\Gamma^r\oplus\Gamma^r}{2}\begin{pmatrix}b^{\mathbf{I}}, & b^{\mathbf{II}}, & b^{\mathbf{III}}, & b^{\mathbf{IV}}, & b^{\mathbf{V}}, &b^{\mathbf{VI}}, & b^{\mathbf{VII}}, & b^{\mathbf{VIII}}\end{pmatrix},\\
                M_{response}&=\frac{1-\Gamma^r\oplus\Gamma^r}{2}\begin{pmatrix}a^{\mathbf{I}}, & a^{\mathbf{II}}, & a^{\mathbf{III}}, & a^{\mathbf{IV}}, & a^{\mathbf{V}}, &a^{\mathbf{VI}}, & a^{\mathbf{VII}}, & a^{\mathbf{VIII}}\end{pmatrix}\\
                &-\frac{1+\Gamma^r\oplus\Gamma^r}{2}\begin{pmatrix}b^{\mathbf{I}}, & b^{\mathbf{II}}, & b^{\mathbf{III}}, & b^{\mathbf{IV}}, & b^{\mathbf{V}}, &b^{\mathbf{VI}}, & b^{\mathbf{VII}}, & b^{\mathbf{VIII}}\end{pmatrix},\\
                G(\omega,k)&=i\Gamma^0M_{response}{M_{source}}^{-1},
            \end{aligned}
        \end{equation}
        where we use Roman numerals to denote the sets of $a,b$ which correspond to different IR boundary conditions for \eqref{CoexistProbeFermionEquationOfMotion}. After obtaining Green's function for probe fermions, the topological Hamiltonian for holographic coexisting semimetals can be defined from the zero frequency Green's function \cite{TopologicalHamiltonianwang-prx1, TopologicalHamiltonianwang-prx2, TopologicalHamiltonianWang:2012ig}
        \begin{equation}
            \label{TopologicalHamiltonian}
            H(k)=-G^{-1}(0,k).
        \end{equation}
        
        With the topological Hamiltonian established, we may now compute the corresponding topological invariants by applying the same computing protocol developed for weak coupling systems. In the following we will calculate the topological invariants for holographic Weyl-nodal line semimetals in Weyl-Nodal phase, from whose IR geometry the IR boundary condition for \eqref{CoexistProbeFermionEquationOfMotion} can be obtained. The IR geometry of the background geometry for this phase is \cite{HolographicWeylNodalChu_2024}
        \begin{equation}
            \label{CoexistWeylNodalIRGeometry}
            \begin{aligned}
                u&=u_0 r^2(1+\delta u r^{\alpha_1}),\\
                f&=f_0 r^\alpha(1+\delta f r^{\alpha_1}),\\
                h&=r^2(1+\delta h r^{\alpha_1}),\\
                A_z&=a_0+\exp\left(-\frac{3 a_0}{r\sqrt{u_0}}\right)r^{\alpha-1},\\
                B_{xy}&=r^\alpha(1+\delta B_{xy} r^{\alpha_1}),\\
                B_{tz}&=B_{tz0} r^2(1+\delta B_{tz} r^{\alpha_1}),\\
                \phi_1&=\phi_{10}\exp\left(-\frac{3 a_0}{2r\sqrt{u_0}}\right)r^{-\frac{\alpha+1}{2}},\\
                \phi_2&=\phi_{20} r^\beta,
            \end{aligned}
        \end{equation}
        where $\delta_u,a_0,\phi_{10},\phi_{20}$ are shooting parameters. Other parameters are $u_0=2.727, f_0=0.635, B_{tz0}=0.787, \alpha=0.183, \alpha_1=1.273, \beta=0.228, \delta f=-2.616\delta u, \delta h=\delta u, \delta B_{xy}=-0.302\delta u, \delta B_{tz}=1.719\delta u$, which are fixed from the parameters in the action. Under this IR geometry, the near horizon solution for probe fermions is
        \begin{equation}
            \label{ProbeFermionCoexistWeylNodalNearHorizonSolution}
            \begin{aligned}
                \psi_1&=\left\{\left[\frac{1}{\sqrt{r}}K_{m+\frac{1}{2}}\left(\frac{{k}_1}{r}\right)+\frac{ik_{1\mu}\Gamma^\mu}{|k_1|}\frac{1}{\sqrt{r}}K_{m-\frac{1}{2}}\left(\frac{{k}_{1}}{r}\right)\right]\right.\\
                &\oplus\left. e^{i\frac{k}{r}}\left(1-\frac{k_\mu\Gamma^\mu}{k}\right)\right\}C_1,~C_1\in \Im\frac{1+\Gamma^r\oplus\Gamma^r}{2},\\
                \psi_2&=\left\{\left[\frac{1}{\sqrt{r}}K_{m+\frac{1}{2}}\left(\frac{{k}_2}{r}\right)-\frac{ik_{2\mu}\Gamma^\mu}{|{k}_2|}\frac{1}{\sqrt{r}}K_{m-\frac{1}{2}}\left(\frac{{k}_2}{r}\right)\right]\right.\\
                &\oplus\left. e^{i\frac{k}{r}}\left(1+\frac{k_\mu\Gamma^\mu}{k}\right)\right\}C_2,~C_2\in \Im\frac{1-\Gamma^r\oplus\Gamma^r}{2},
            \end{aligned}
        \end{equation}
        where $k_\mu=(-\frac{\omega}{u_0},0,0,\frac{k_z}{\sqrt{u_0}}), k_{1\mu}=(-\frac{\omega}{u_0},0,0,\frac{k_z-q a_0}{\sqrt{u_0}}), k_{2\mu}=(-\frac{\omega}{u_0},0,0,\frac{k_z+q a_0}{\sqrt{u_0}}), k=|\mathbf{k}|$. Details for the derivation of this near horizon solution can be {found in} appendix \ref{ProbeFermionNearHorizonSolution}. 
        Using \eqref{CoexistWeylNodalIRGeometry} as IR boundary conditions to get the background field $u,f,h,A_z,B_{xy},B_{tz},\phi_1,\phi_2$ numerically, the equations  \eqref{CoexistProbeFermionEquationOfMotion} can be solved numerically with \eqref{ProbeFermionCoexistWeylNodalNearHorizonSolution} as {IR boundary conditions}

\begin{figure}[htbp]
            \centering
            \includegraphics[width=0.8\textwidth]{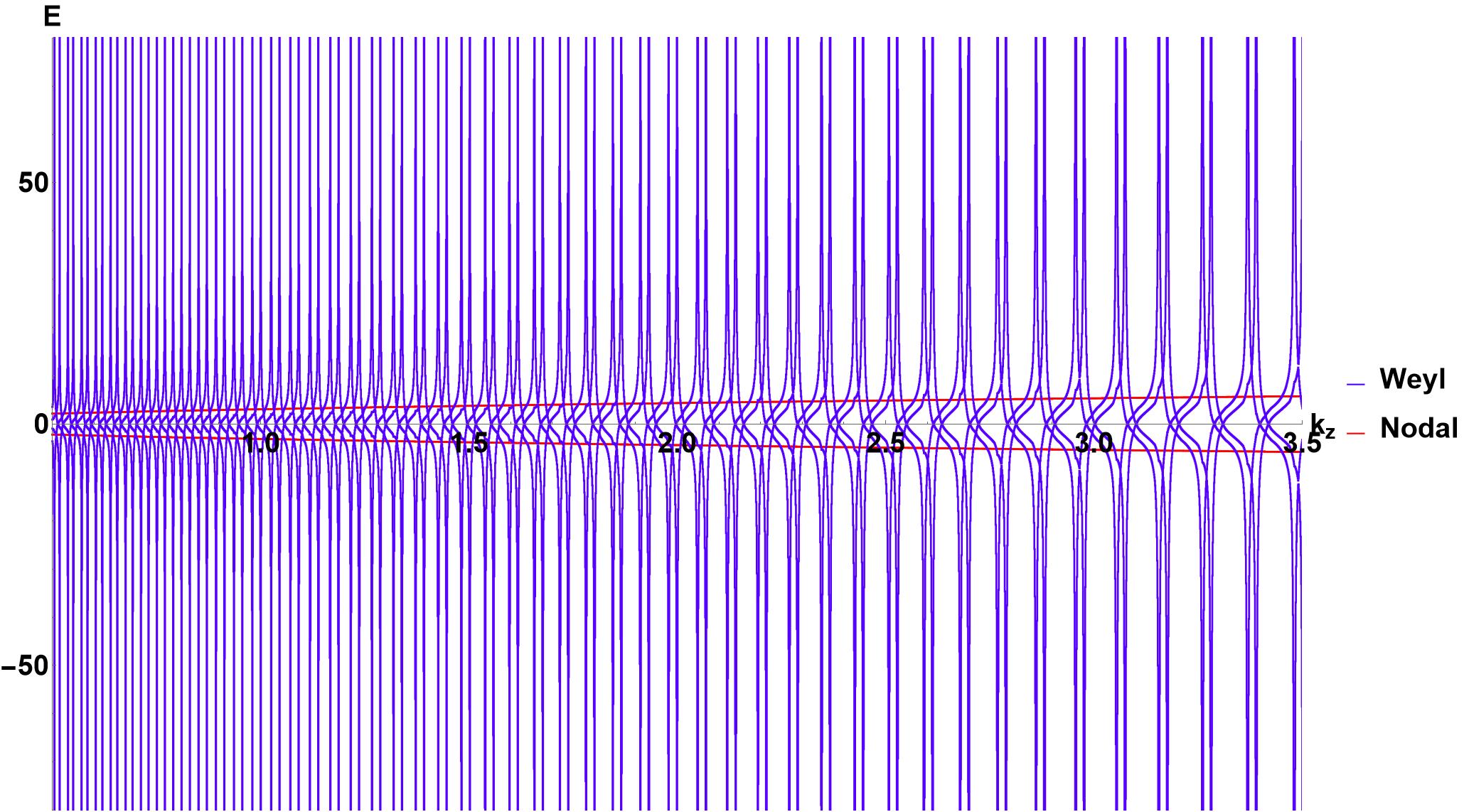}
            \caption{The effective band structure from the topological Hamiltonian in the Weyl-Nodal phase with $M_1/b=0.701,M_2/c=0.807$ along the $k_z$ axis. {The bands from  the Weyl sector are plotted in blue and bands from the nodal line sector plotted in red.}}
            \label{FigureWeylNodalKz}
        \end{figure}

\begin{figure}[htbp]
            \centering
            \includegraphics[width=0.8\textwidth]{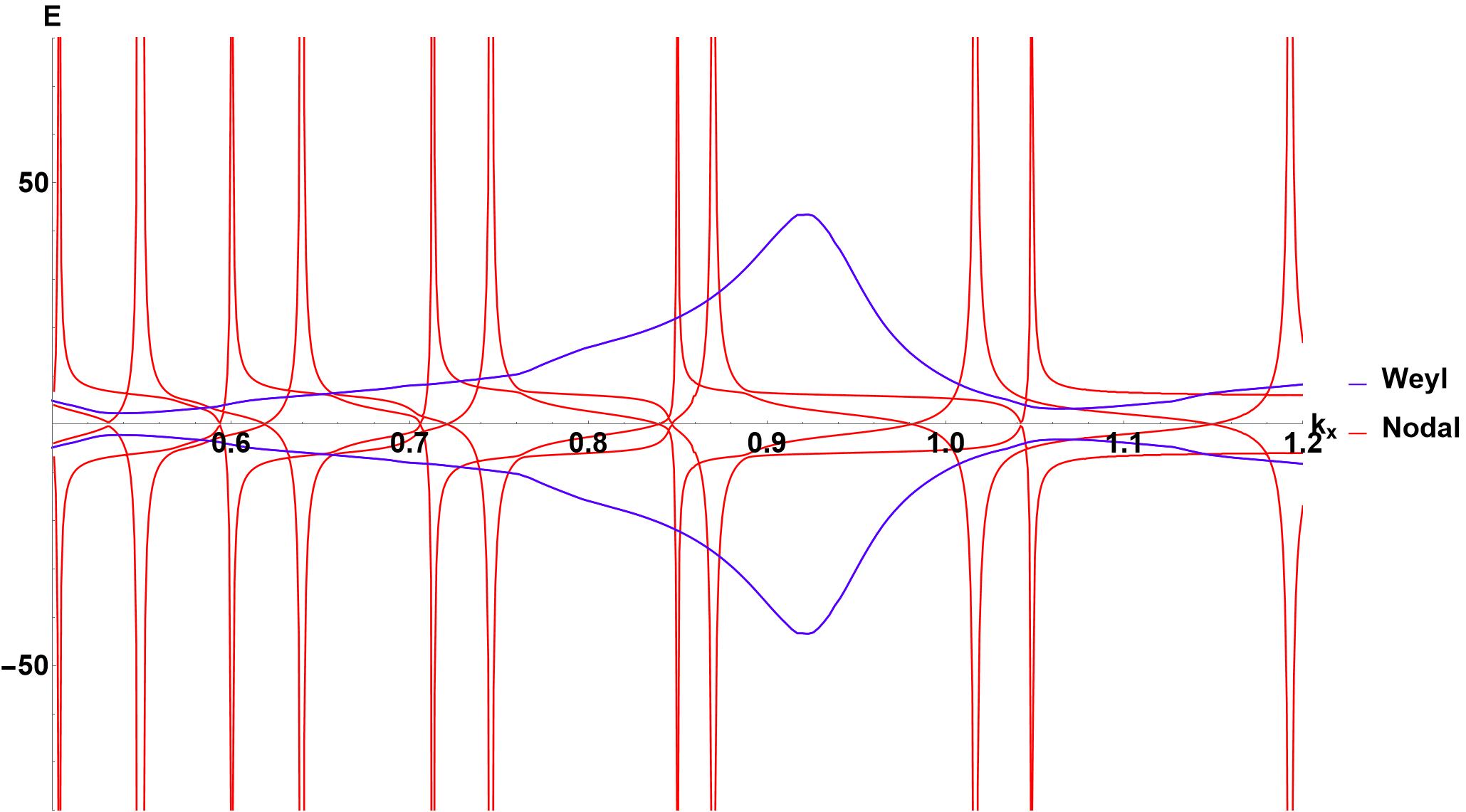}
            \caption{The band structure for the topological Hamiltonian in the Weyl-Nodal phase with $M_1/b=0.701,M_2/c=0.807$ along the $k_x$ axis. {The Weyl sector is plotted in blue and the nodal line sector in red.}}
            \label{FigureWeylNodalKx}
        \end{figure}

      \noindent {\bf Effective topological band structures of the holographic coexisting semimetal. }  
        {We select eight sets of linearly independent IR boundary conditions, from which we derive the corresponding sources and responses. 
    By applying the source-response relations \eqref{CoexistProbeFermionGreenFunction}, we obtain the associated Green's functions, and finally construct the topological Hamiltonian for the Weyl-Nodal phase using its definition.}
        After getting the topological Hamiltonian for the Weyl-Nodal phase, we can plot the effective topological band structures indicated by eigenvalues of the topological Hamiltonian along the $k_z$ axis in Fig. \ref{FigureWeylNodalKz}, where Weyl nodes reside, and the eigenvalues along the $k_x$ axis in Fig. \ref{FigureWeylNodalKx}, which intersects the nodal line. 

{By construction, due to the ansatz \eqref{BackGroundZeroTemperatureAnsatz} for the background fields, Weyl nodes should emerge in pairs along the $k_z$ axis, while the nodal line is expected to lie within the $k_z=0$ plane. Conversely, the Weyl sector remains gapped in the $k_z=0$ plane and the nodal line sector should be gapped along the $k_z$ direction.}
        
        
        As it is shown in Fig. \ref{FigureWeylNodalKz}, along the $k_z$ axis, these effective energy bands are colored with red and blue separately to distinguish contributions from the nodal-line sector and Weyl sector, respectively. Notably, the bands contributed by the nodal line sector exhibit a full band gap while maintaining twofold degeneracy and relatively flat dispersion throughout the $k_z$ axis. In contrast, the Weyl semimetal contribution manifests as two sets of similarly shaped bands that form multiple Fermi surface structures. The Weyl points appear in pairs, with both the intra-pair separation and inter-pair spacing increasing monotonically with $k_z$. Particularly intriguing is the observed band crossing ordering interchange between the two sets of Weyl bands along the $k_z$ axis, as evidenced by their topological charges. This phenomenon is consistent with our previous {findings}\cite{HolographicWeylZ2chen2025} {that the bandcrossing orders for two sets of bands may interchange}.

        As shown in Fig. \ref{FigureWeylNodalKx}, along the $k_x$ axis, we employ the same color scheme to distinguish bands coming from different sectors. In this direction, we can observe twofold degenerate nodes (with eigenvalue zero) originating from the nodal-line semimetal. The bands from the nodal line semimetal sector consist of two distinct sets with markedly different shapes, forming multiple Fermi surfaces. Notably, a band crossing ordering interchange between these two sets can be clearly observed from their dispersion, a phenomenon not reported in previous studies. Similar to the behavior along the $k_z$ axis, both the intra-pair and inter-pair spacings of the Fermi surfaces increase with $k_x$.

        In contrast, the bands from the Weyl sector along the $k_x$ axis remain fully gaped and twofold degenerate. However, unlike the bands from the nodal-line sector along the $k_z$ axis, which exhibit a relatively flat dispersion, the Weyl bands here display a near-periodic oscillatory upward trend with increasing $k_x$. {Along the $k_x$ direction, the nodal points of the topological Hamiltonian exhibit rapidly increasing separation with growing $k_x$. To ensure clear visualization of the topological defect distribution, we adopt distinct plotting ranges for the $k_x$ and $k_z$ directions: the $k_x$ range is optimally selected to maintain nodal point resolution.}

        \noindent  {\bf Topological invariants of the holographic coexisting semimetal.}
        Owing to the system's high symmetry, we systematically investigate the stability of topological defects under various perturbations by computing the following topological invariants:
        \begin{enumerate}
            \item Stability under mirror-symmetric perturbations: The $\zeta_0$ and $\widetilde{\zeta}_2$ invariants are computed to examine the stability of the nodal rings against perturbations that preserve mirror symmetry.
            \item Stability under general perturbations: The $\zeta_1$ and $\zeta_2$ invariants are calculated to determine the stability of the nodal rings against arbitrary general perturbations.
            \item Stability of Weyl nodes: The Weyl charges are computed to analyze the stability of the Weyl nodes.
        \end{enumerate}

        The specific computational schemes are as follows:
        \begin{enumerate}
            \item  Weyl charge: For each Weyl node, choose an enclosing spherical surface and integrate the Berry curvature of the topological Hamiltonian over this surface.
            \item $\zeta_0$ invariant: For a given nodal ring, select one point inside and one outside the ring, and compare the eigenvalues of the mirror operator at these two points.
            \item $\widetilde{\zeta}_2$ invariant: For the same nodal ring, select one closed integration path inside and one outside the ring, and compare the average phases of the Wilson loops on these paths.
            \item $\zeta_1$ invariant: For a nodal ring, choose an integration loop encircling it and compute the line integral of the Berry connection along this loop.
            \item $\zeta_2$ invariant: For a nodal ring, construct a torus enclosing it and compute the winding number of the Wannier centers on this torus.
        \end{enumerate}
        
        {The final results are the following: (1) the Weyl charges of the Weyl nodes are quantized to be $\pm 1$\cite{HolographicWeylZ2chen2025}, and (2) the topological invariant $\zeta_1(\zeta_0)$ takes the value of 1 for the nodal ring in Weyl-nodal line coexisting semimetals.\cite{HolographicNodalLineSemimetalsLiu_2021}.} (3) The $\zeta_2(\widetilde{\zeta_2})$ invariant takes the value of $0$ for the nodal ring in Weyl-nodal line coexisting semimetals. In summary, numerical results ensure that the holographic model \eqref{CoexistHolographicModel} with the IR geometry \eqref{CoexistWeylNodalIRGeometry} truly describes coexisting semimetals equipped with both topological protected Weyl nodes and a nodal ring.
\section{Topological invariants and effective band structure for critical phases in holographic semimetals}
    \label{Section4}
    {As previously stated, a non-zero topological invariant is indicative of a topological non-trivial phase, whilst a zero topological invariant signifies a topological trivial phase, corresponding to a phase wherein {no band touching points or only topologically trivial band crossings} exist within the energy band structure. It is therefore both interesting and vital to investigate the value of the topological invariant of the critical phase of these semimetal states to further confirm the critical nature of the phases. 
    In the critical phases of Weyl or nodal line semimetals, the Weyl nodes merge to form a topologically trivial Dirac node or the nodal ring shrinks to a trivial Dirac node.}

    In holographic semimetals, critical phases have been less studied, especially their band structures and topological invariants.  
    The IR geometry corresponding to the critical phases in holographic semimetals is described by a Lifshitz metric. This configuration exhibits remarkable stability, manifesting as the critical phase across all possible shooting parameter selections.
    The critical phases in holographic semimetals should feature band-touching nodes with vanishing topological charge, ensuring consistency with the weak-coupling limit scenario.

    In this section we will investigate the behavior of probe fermions in critical phases of holographic topological semimetals, especially focusing on the location of critical nodes, the effective band structures and topological invariants.  
    We will calculate the topological Hamiltonian and plot the corresponding effective band structures for the pure holographic Weyl semimetal (case I), the pure holographic nodal line semimetal (case II), and the Weyl-Critical phase of the holographic coexisting semimetal (case III), respectively. {We emphasize that this study focuses exclusively on the Weyl-Nodal and Weyl-Critical phases, while deliberately omitting analysis of other configurations (e.g., Critical-Nodal or Critical-Critical phases) for two compelling reasons: (1) The established understanding of Weyl-Nodal and Weyl-Critical phases enables reliable extrapolation to predict the essential physics of other phases; (2) To maintain conceptual clarity and avoid complicated numerics for certain unstable systems or unnecessary digressions.}
    
    \subsection{Case I: The holographic Weyl semimetal at the critical phase}
        To begin with, we first study the critical phase for the pure holographic Weyl semimetal. At the critical phase,  the two Weyl nodes merge and form a Dirac node. The holographic model for Weyl semimetals \eqref{WeylHolographicModel} has been introduced in section \ref{Section2}. Here to get the background geometry for the critial phase solution, we need to give the IR geometry for the critical phase. 
        With the ansatz $g=\frac{dr^2}{u^2}+u^2 dx^2+u^2 dy^2+h^2 dz^2, A=A_zdz,\Phi=\phi$, where $u,h,A_z,\phi$ are all real functions for $r$, we have the following IR geometry for the critical phase of the Weyl semimetal \cite{HolographicWeylSemimetalsLandsteiner_2016}
        \begin{equation}
            \label{PureWeylCriticalIRGeometry}
            \begin{aligned}
                u&=u_0 r^2(1+\delta u r^\alpha),\\
                h&=h_0 r^{\beta}(1+\delta h r^\alpha),\\
                A_z&=r^\beta(1+\delta a r^\alpha),\\
                \phi&=\phi_0(1+\delta\phi r^\alpha),
            \end{aligned}
        \end{equation}
        with the parameters taking the values $\alpha=1.135,\beta=0.407,u_0=1.468,h_0=0.344,\phi_0=-.947,\delta u=0.369\delta\phi,\delta h=-2.797\delta\phi,\delta a=0.137\delta\phi$, when we set the following values for the parameters in \eqref{WeylHolographicModel}: $q=1,m^2=-3,\lambda=\frac{1}{10},\alpha=1,2\kappa^2=L=1$.
        When the shooting parameter $\delta\phi$ takes the value $-1$, the background geometry for the critical phase could be obtained by integrating the system to the UV boundary.
        
\begin{figure}[htbp]
            \centering        \includegraphics[width=0.8\textwidth]{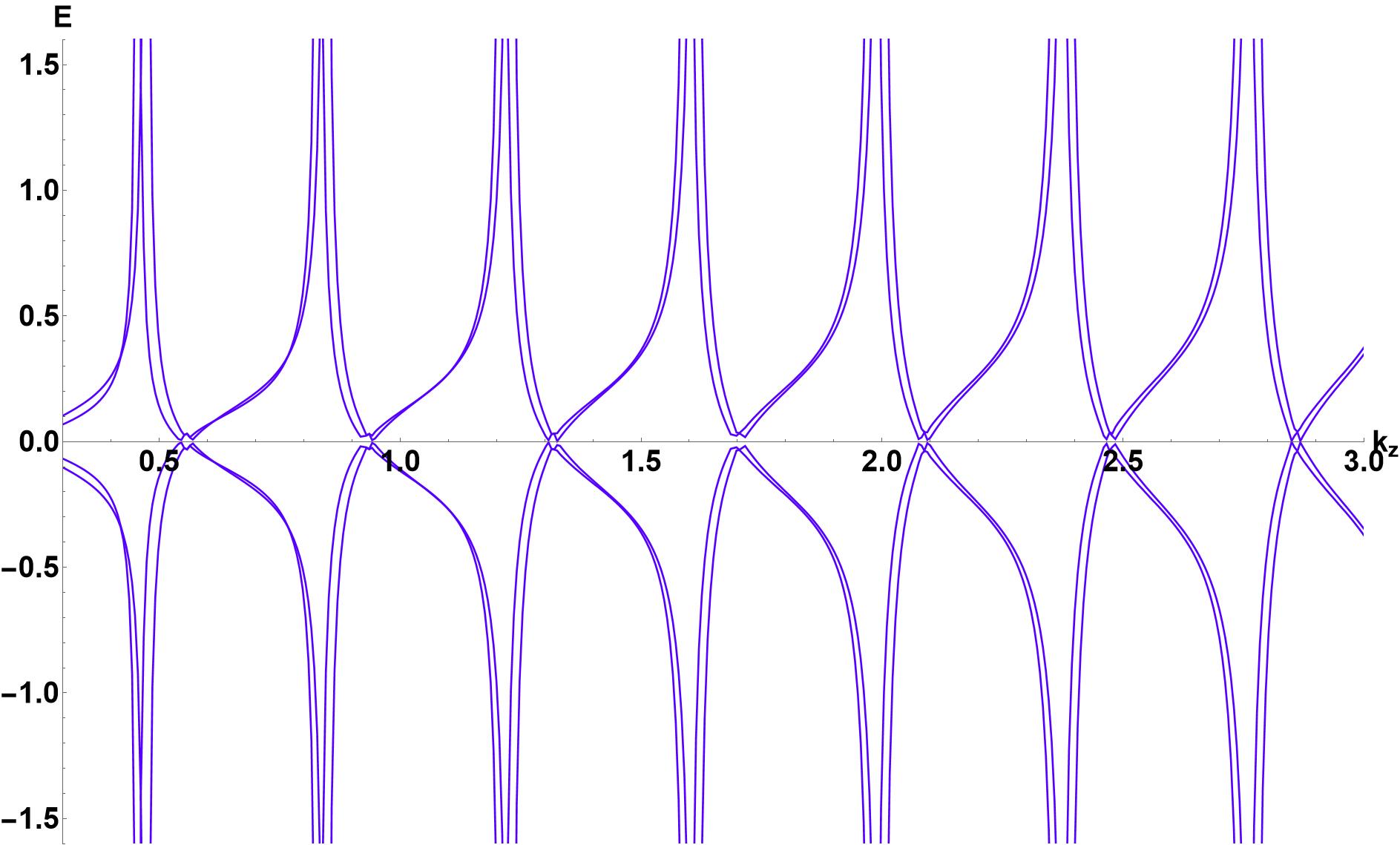}
            \caption{The band structure for the topological Hamiltonian corresponding to the Critical phase for Weyl semimetals with $M/b=0.8597$ along axis $k_z$. Note that the numerical results exhibit slight deviations from the ideal results due to inherent computational challenges. At the Fermi surface, where the topological Hamiltonian $H(k)$ becomes nearly a zero matrix (reflecting the expected fourfold degeneracy), the relation $H(k)=-G^{-1}(\omega=0,k)$ forces the Green's function $G$ to be highly singular. This numerical instability leads to small but observable errors in the band structure.}
            \label{FigurePureWeylCriticalKz}
        \end{figure}

        The bulk action for probe fermions \eqref{WeylProbeFermionBulkAction} in this system has  been introduced in section \ref{Section3}. 
        We have the corresponding near horizon solutions for probe fermions from appendix \ref{ProbeFermionNearHorizonSolution}
        \begin{equation}
            \label{ProbeFermionPureWeylCriticalNearHorizonSolution}
            \begin{aligned}
                \psi_1&=e^{i\frac{k}{r}}\left(1-\frac{k_\mu\Gamma^\mu}{k}\right)C_1,~C_1\in \Im\frac{1+\Gamma^r}{2},\\
                \psi_2&=e^{i\frac{k}{r}}\left(1+\frac{k_\mu\Gamma^\mu}{k}\right)C_2,~C_2\in \Im\frac{1-\Gamma^r}{2},
            \end{aligned}
        \end{equation}
        where $k^\mu=(-\frac{\omega}{u_0},\frac{k_x}{u_0},\frac{k_y}{u_0},0),k=|\mathbf{k}|$. Then we can use the procedure that has been introduced in section \ref{Section3} to get the topological Hamiltonian. After that we can plot the effective topological band structure for the holographic Weyl semimetal in the critical phase along the $k_z$ axis, as shown in Fig. \ref{FigurePureWeylCriticalKz}, where all the Dirac nodes {locate}. {Note that in Fig. \ref{FigurePureWeylCriticalKz} the numerical results exhibit slight deviations from the ideal results due to inherent computational challenges. At the Fermi surface, where the topological Hamiltonian $H(k)$ becomes nearly a zero matrix (reflecting the expected fourfold degeneracy), the relation $H(k)=-G^{-1}(\omega=0,k)$ forces the Green's function $G$ to be highly singular. This numerical instability leads to small but observable errors in the band structure.}
        
        In the critical phase, the effective band structure of the holographic Weyl semimetal closely resembles that of the topologically non-trivial phase, except that two nearly degenerate bands merge, resulting in an overall twofold-degenerate dispersion. As the separation between a pair of Weyl points approaches zero in the critical phase, they coalesce into a fourfold degenerate Dirac point. This behavior has been explicitly verified through calculations of the topological invariants. We employ the previously introduced methods for the Weyl charge to evaluate the topological charge of Dirac nodes in the Critical phase. Specifically, we compute the surface integral of Berry curvature over the sphere enclosing the Dirac nodes,  and obtain the final result, which is $0$, confirming the topological properties of these critical Dirac nodes.
        
        We now explain the origin of the persistent twofold band degeneracy observed in the critical-phase Weyl semimetal along the $k_z$-axis direction through investigating the equation of motion for probe fermions in the near horizon region, because the nature of the topological Hamiltonian is mainly governed by the near horizon region. {According} to (\ref{WeylProbeFermionBulkAction}), the equation of motion expanded to the leading and subleading orders are
        \begin{equation}
            \begin{aligned}
                \left(\sqrt{u_0}r\Gamma^r\partial_r+i\frac{k_x}{\sqrt{u_0}r}\Gamma^x+i\frac{k_y}{\sqrt{u_0}r}\Gamma^y-m\right)\psi_1-\eta\phi_0\psi_2=0,\\
                \left(\sqrt{u_0}r\Gamma^r\partial_r+i\frac{k_x}{\sqrt{u_0}r}\Gamma^x+i\frac{k_y}{\sqrt{u_0}r}\Gamma^y+m\right)\psi_2-\eta\phi_0\psi_1=0.
            \end{aligned}
        \end{equation}
        
        When we calculate the topological Hamiltonian along the $k_z$ axis at $k_x=k_y=0$, the first and second components of $\psi_1$ share the same equation of motion, and the third and fourth components of $\psi_1$ share the same equation of motion. The same holds for $\psi_2$. Thus it is reasonable that the four bands for topological Hamiltonian are pairwise degenerate. {However, due to non-negligible contributions from regions beyond the near horizon zone, the resulting band structure exhibits approximate degeneracy rather than exact degeneracy. Later we will demonstrate that the critical phase for holographic nodal line semimetal exhibits fundamentally distinct behavior: the band degeneracy occurs exclusively at the crossing points, while non-degenerate bands persist throughout the remainder of the momentum space.}
    \subsection{Case II: The holographic nodal line semimetal at the critical phase}
\begin{figure}[htbp]
            \centering
            \includegraphics[width=0.8\textwidth]{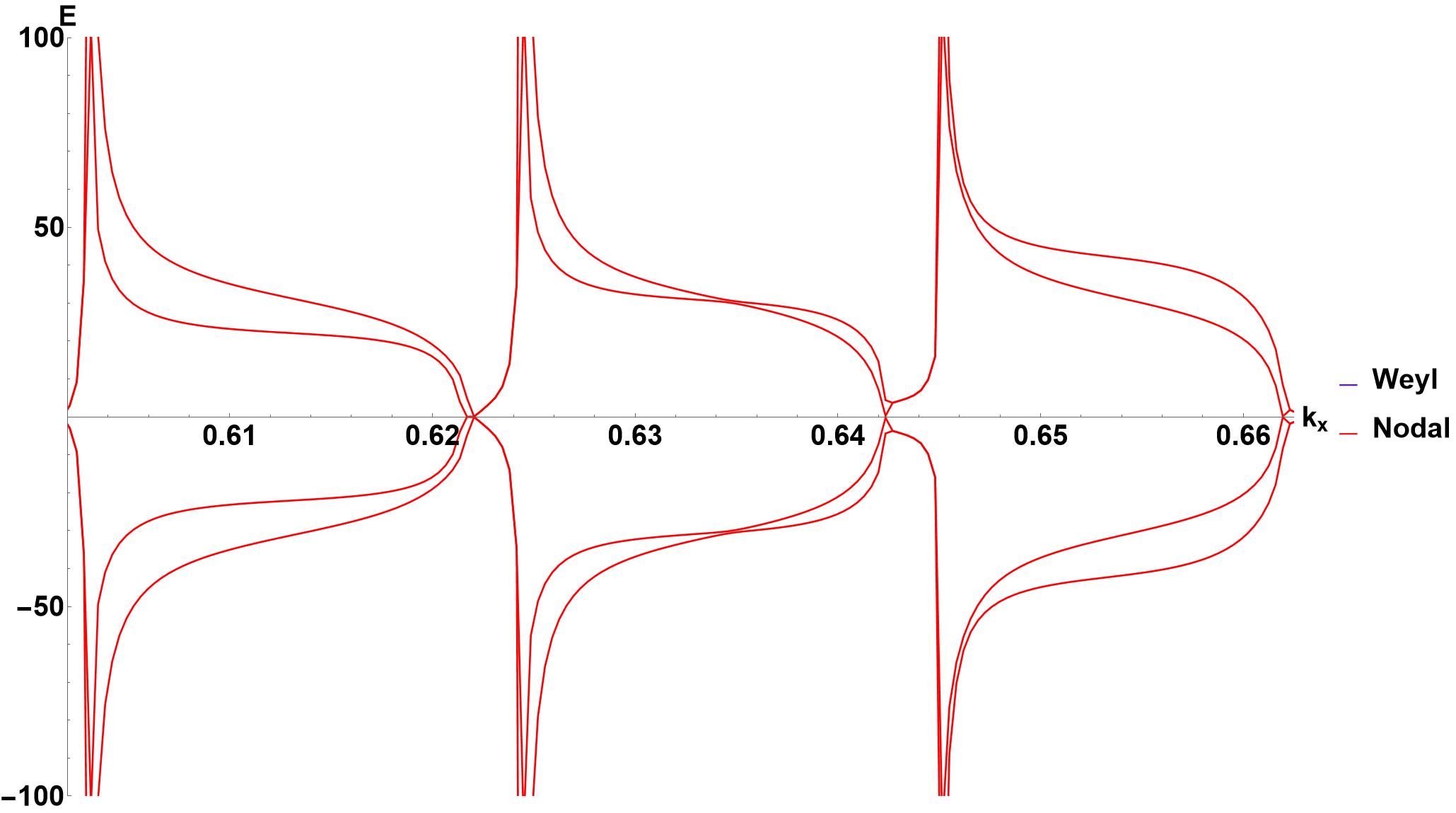}
            \caption{The band structure for the topological Hamiltonian corresponding to the Critical phase for nodal line semimetals with $M/c=0.859$ along axis $k_x$ under the Critical situation.}
            \label{FigurePureNodalCriticalKx}
        \end{figure}

        At the critical phase in the nodal line semimetal, the nodal line shrinks to a critical point. The critical point will be gaped under small perturbations if the corresponding $\zeta_2$ invariant is $0$. On the contrary, the critical point will expand back into a nodal ring if $\zeta_2=1$. The holographic model for the nodal semimetal \eqref{NodalHolographicModel} has been presented   and the corresponding action for probe fermions is \eqref{NodalProbeFermionBulkAction}. Here we takes the ansatz $g=\frac{dt^2}{u^2}+f dx^2+f dy^2+u dz^2, B=\frac{1}{2}(B_{xy}dx\wedge dy+iB_{tz}dt\wedge dz), \Phi=\phi$ where $u,f,B_{xy},B_{tz},\phi$ are real functions for $r$, and the IR geometry of the critical phase for the holographic nodal line semimetal is\cite{HolographicNodalLineSemimetalsLiu_2021}
        \begin{equation}
            \label{PureNodalCriticalIRGeometry}
            \begin{aligned}
                u&=u_0 r^2(1+\delta u r^\alpha),\\
                f&=f_0 r^{\beta}(1+\delta f r^\alpha),\\
                B_{tz}&=b_{tz0}r^2(1+\delta b_{tz} r^\alpha),\\
                B_{xy}&=b_{xy0}r^\beta(1+\delta b_{xy} r^\alpha),\\
                \phi&=\phi_0(1+\delta\phi r^\alpha),
            \end{aligned}
        \end{equation}
        where $\alpha=1.274,\beta=0.314$, and $u_0=2.735,f_0=0.754 b_{xy0},\phi_0=0.557,b_{tz0}=1.437, \delta u=0.882\delta\phi, \delta f=-2.151\delta\phi, \delta b_{tz}=1.718\delta\phi, \delta b_{xy}=-0.254\delta\phi$ with $\delta\phi,b_{xy0}$ being the shooting parameters. 
        The Lifshitz type symmetry makes sure that we can set $\delta\phi=-1$ and $b_{xy0}=1$ without loss of generality. 
        With the IR geometry for critical phase, the near horizon solution for probe fermions can be obtained
        \begin{equation}
            \label{ProbeFermionPureNodalCriticalNearHorizonSolution}
            \begin{aligned}
                \psi_1&=e^{i\frac{k}{r}}\left(1-\frac{k_\mu\Gamma^\mu}{k}\right)C_1,~C_1\in \Im\frac{1+\Gamma^r}{2},\\
                \psi_2&=e^{i\frac{k}{r}}\left(1+\frac{k_\mu\Gamma^\mu}{k}\right)C_2,~C_2\in \Im\frac{1-\Gamma^r}{2},
            \end{aligned}
        \end{equation}
        where $k^\mu=(-\frac{\omega}{u_0},0,0,\frac{k_z}{u_0}),k=|\mathbf{k}|$. Then we can use the procedure which has been introduced in section \ref{Section3} to get the topological Hamiltonian. Consequently, the effective band structure for the holographic nodal line semimetal in the critical phase could be obtained from the topological Hamiltonian. The band structure along the $k_x$ axis is plotted\footnote{Due to the same numerical limitations, the results presented in this figure exhibit similar non-ideal artifacts as those observed in Fig. \ref{FigurePureWeylCriticalKz}.} in Fig. \ref{FigurePureNodalCriticalKx}.
        
        In the critical phase, the band structure of the nodal-line semimetal exhibits striking differences from that of the critical-phase Weyl semimetal. Unlike the Weyl system where bands maintain global twofold degeneracy, the effective bands in the nodal line semimetal remain non-degenerate everywhere except at the critical point, where they form a fourfold degeneracy. This behavior closely resembles the nodal-line semimetal in its topologically non-trivial phase, {in which} the separation between a pair of degenerate points vanishes. Below we systematically analyze the origin of this phenomenon. Using the same method in Weyl semimetals, we will investigate the equation of motion in the near horizon region with $k_z=0$.
        \begin{equation}
            \begin{aligned}
                \left(\sqrt{u_0}r\Gamma^r\partial_r-m\right)\psi_1-\eta_1\phi_0\psi_2+2\eta_2\begin{pmatrix}b_{tz0}+b_{xy0}& & & \\ & -b_{tz0}-b_{xy0} & & \\ & & b_{tz0}-b_{xy0} & \\ & & & -b_{tz0}+b_{xy0}\end{pmatrix}\psi_2=0,\\
                \left(\sqrt{u_0}r\Gamma^r\partial_r+m\right)\psi_2-\eta_1\phi_0\psi_1+2\eta_2\begin{pmatrix}b_{tz0}-b_{xy0}& & & \\ & -b_{tz0}+b_{xy0} & & \\ & & b_{tz0}+b_{xy0} & \\ & & & -b_{tz0}-b_{xy0}\end{pmatrix}\psi_2=0.
            \end{aligned}
        \end{equation}
        
        The equations of motion for all the components of $\psi_1$ and $\psi_2$ are different, so it is natural that the four bands for nodal line semimetals in the critical phase are not degenerate except at the critical point. {Notably, in the critical phase, the holographic nodal line semimetal exhibits behavior distinct from the weak coupling case: instead of contracting to a point, a pair of nodal lines coincide into a single nodal ring. Although the $\zeta_1$ invariant is a mod 2 invariant ($\mathbb{Z}_2$ topological number), it can vanish when two nodal rings merge. Through direct integration of the Berry connection, we confirm this phenomenon: the holographic nodal ring in the critical phase indeed exhibits a vanishing $\zeta_1$ invariant.}
    \subsection{Case III: The holographic coexisting semimetal at the Weyl-Critical phase}
        To investigate the effective band structure for the Weyl-Critical phase in the holographic coexisting semimetal, we will start from the IR geometry  for \eqref{CoexistHolographicModel} that gives the whole background geometry of the Weyl-Critical phase after integration as follow\cite{HolographicWeylNodalChu_2024}
        \begin{equation}
            \label{CoexistWeylCriticalIRGeometry}
            \begin{aligned}
                u&=u_0 r^2(1+\delta u r^{\alpha_1}),\\
                f&=f_0 r^{2\alpha}(1+\delta f r^{\alpha_1}),\\
                h&=h_0 r^2(1+\delta h r^{\alpha_1}),\\
                A_z&=a_0+{\phi_{10}}^2 h_0 \exp\left(-\frac{3 a_0}{r\sqrt{u_0h_0}}\right)r^{1-\alpha},\\
                B_{xy}&=r^\alpha(1+\delta B_{xy} r^{\alpha_1}),\\
                B_{tz}&=B_{tz0} r^2(1+\delta B_{tz} r^{\alpha_1}),\\
                \phi_1&=\phi_{10}\exp\left(-\frac{3 a_0}{2r\sqrt{u_0h_0}}\right)r^{-\frac{\alpha+1}{2}},\\
                \phi_2&=\phi_{20}(1+\delta\phi_{2}r^{\alpha_1}),
            \end{aligned}
        \end{equation}
        where the parameters takes the value $u_0=2.735,f_0=0.754,\phi_{20}=0.557,\alpha=0.314,\alpha_1=1.274,\delta u=1.399,\delta f=-3.411,\delta h=1.399,\delta B_{xy}=-0.402,\delta B_{tz}=2.723,\delta\phi_2=1.585,B_{tz0}=0.869\sqrt{h_0}$, and $h_0,a_0,\phi_{10}$ are shooting parameters. Using \eqref{CoexistWeylCriticalIRGeometry} as {IR boundary} conditions, we then get the background field $u,f,h,A_z,B_{xy},B_{tz},\phi_1,\phi_2$ numerically.
        
\begin{figure}[htbp]
            \centering
            \includegraphics[width=0.8\textwidth]{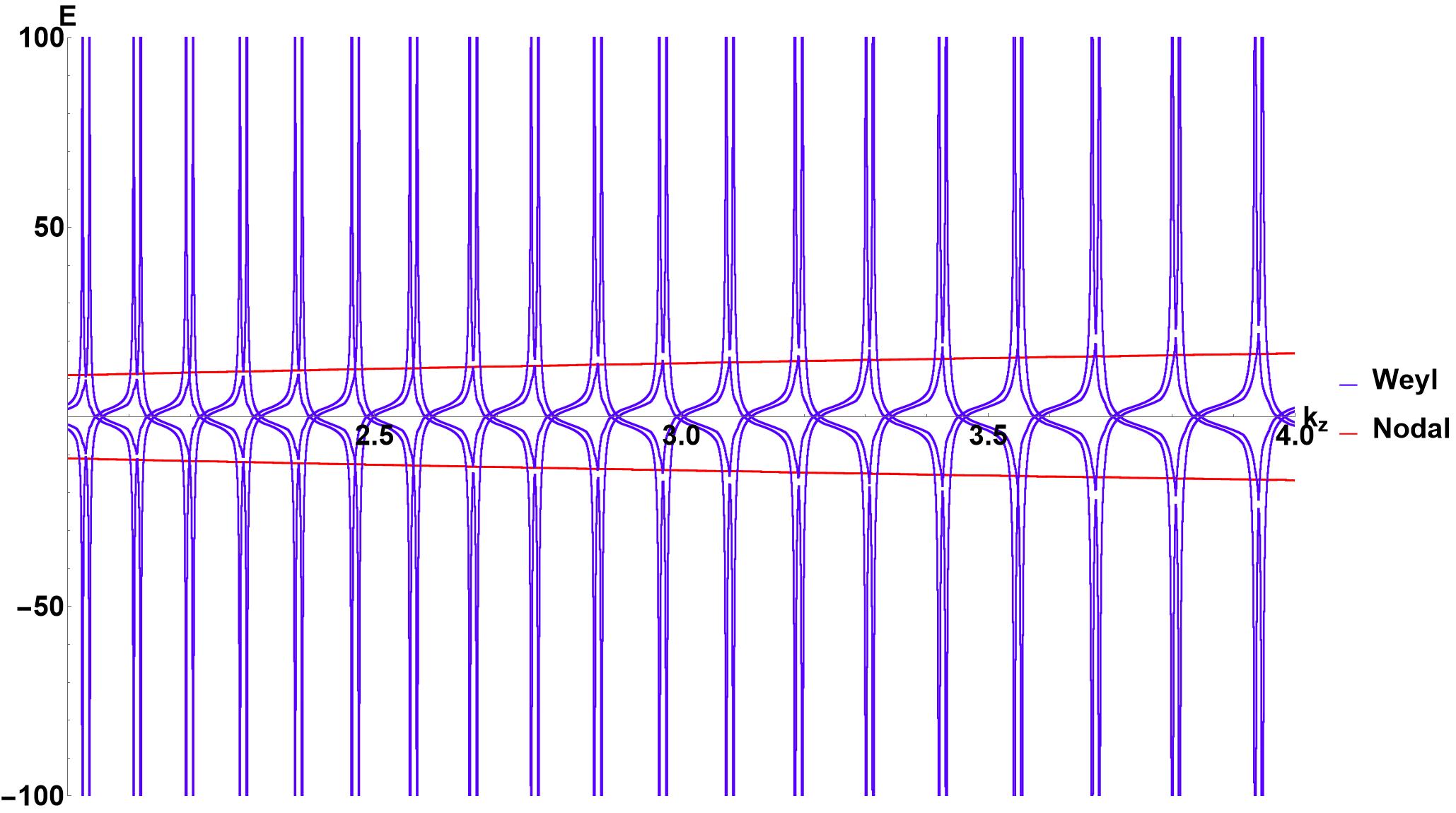}
            \caption{The band structure for the topological Hamiltonian in the Weyl-Critical phase of the holographic  coexisting semimetal with $M_1/b=1.458$ and $M_2/c=0.862$ along the $k_z$ axis. All the Weyl nodes locate on the $k_z$ axis and the nodal line part is gaped along the  $k_z$ axis at $k_x=k_y=0$}
            \label{FigureWeylCriticalKz}
        \end{figure}

\begin{figure}[htbp]
            \centering
            \includegraphics[width=0.8\textwidth]{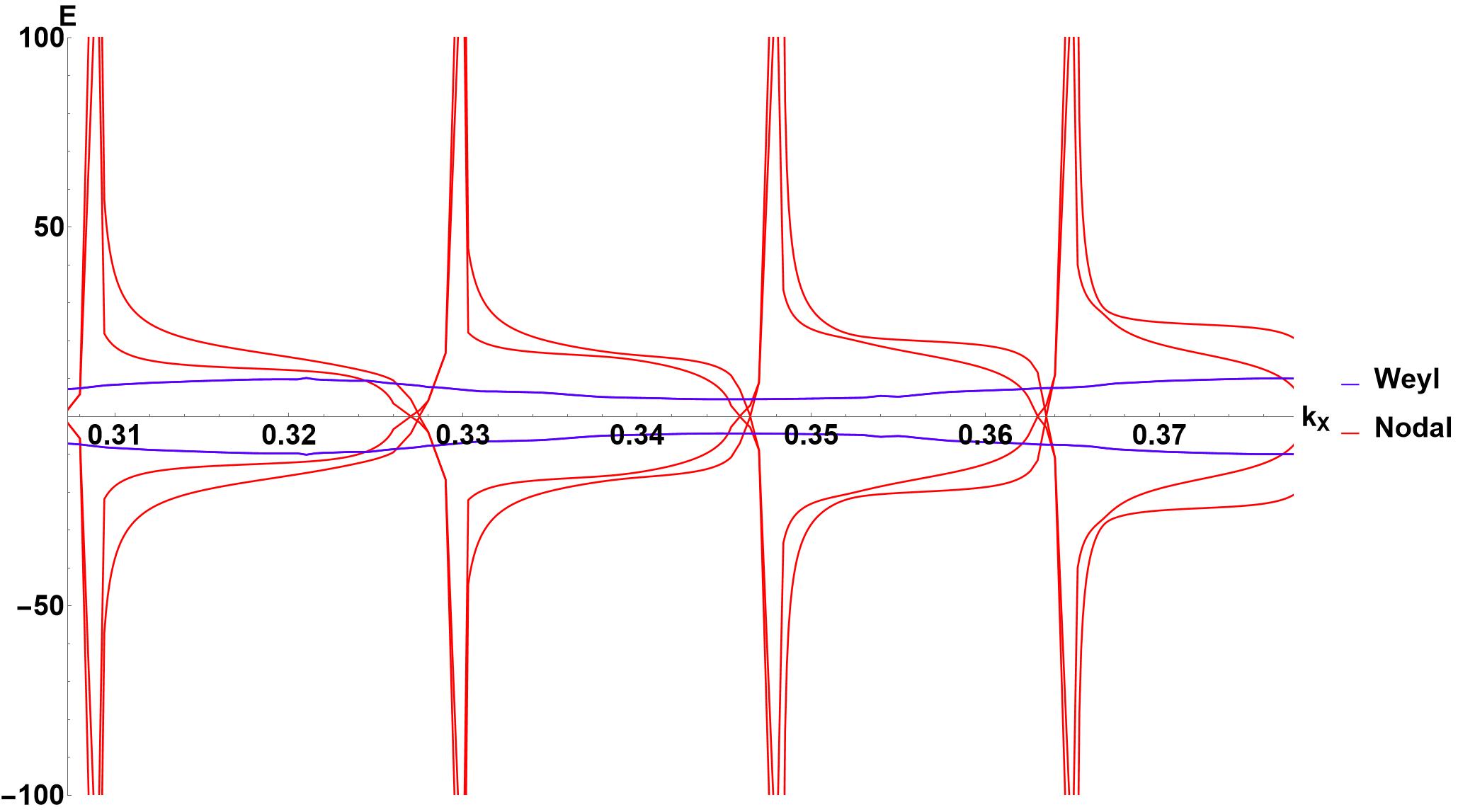}
            \caption{The band structure from the topological Hamiltonian corresponding to the Weyl-Critical phase of the holographic coexisting semimetal with $M_1/b=1.458$ and $M_2/c=0.862$ along the $k_x$ axis at $k_z=0$. The critical nodal rings appear on the $k_z=0$ plane at multiple discrete values of $k=\sqrt{k_x^2+k_y^2}$.  The bands from the Weyl sector are gapped at $k_z=0$}
            \label{FigureWeylCriticalKx}
        \end{figure}

        With the IR geometry \eqref{CoexistWeylCriticalIRGeometry}, the near horizon solution for probe fermions can be {found in} appendix \ref{ProbeFermionNearHorizonSolution}
        \begin{equation}
            \label{ProbeFermionCoexistWeylCriticalNearHorizonSolution}
            \begin{aligned}
                \psi_1&=\left\{e^{-\frac{|k_1|}{r}}\left(1+i\frac{k_{1\mu}\Gamma^\mu}{|k_1|}\right)\oplus e^{i\frac{k}{r}}\left(1-\frac{k_\mu\Gamma^\mu}{k}\right)\right\}C_1,~C_1\in \Im\frac{1+\Gamma^r\oplus\Gamma^r}{2},\\
                \psi_2&=\left\{e^{-\frac{|k_2|}{r}}\left(1-i\frac{k_{2\mu}\Gamma^\mu}{|k_2|}\right)\oplus e^{i\frac{k}{r}}\left(1+\frac{k_\mu\Gamma^\mu}{k}\right)\right\}C_2,~C_2\in \Im\frac{1-\Gamma^r\oplus\Gamma^r}{2},
            \end{aligned}
        \end{equation}
        where $k_\mu=(-\frac{\omega}{u_0},0,0,\frac{k_z}{\sqrt{u_0 h_0}}), k=|\mathbf{k}|, k_{1\mu}=(-\frac{\omega}{u_0},0,0,\frac{k_z-q a_0}{\sqrt{u_0 h_0}}), k_{2\mu}=(-\frac{\omega}{u_0},0,0,\frac{k_z+q a_0}{\sqrt{u_0 h_0}})$. Then the equations of motion for probe fermions \eqref{CoexistProbeFermionEquationOfMotion} can be solved with \eqref{ProbeFermionCoexistWeylCriticalNearHorizonSolution} being {IR boundary} conditions. 
        After obtaining the Green's function numerically, we can get the topological Hamiltonian for the holographic coexisting semimetal in the Weyl-Critical phase and plot the effective band structure along the $k_z$ and the $k_x$ axis\footnote{Note that there exist the same numerical difficulties as for the pure holographic nodal line semimetal in the critical phase, as the Green's function is highly singular near the critical point which leads to numerical instability.} in Fig. \ref{FigureWeylCriticalKz} and Fig. \ref{FigureWeylCriticalKx} separately.
        
        In the picture, we use the color scheme where red denotes  the nodal line sector  and blue represents the Weyl sector. Remarkably, the resulting band structure appears as a combination of the bands of topologically nontrivial phase in the Weyl semimetal bands and the bands in the critical phase of the nodal line semimetal. 
        Through explicit calculation of the relevant topological invariants, we have confirmed that this indeed corresponds to the Weyl-Critical phase, whose Weyl charges are $\pm 1$ and $\zeta_1,\zeta_2$ invariants are 0.
        
\begin{figure}[htbp]
            \begin{minipage}{0.49\linewidth}
                \vspace{3pt}
                \centerline{\includegraphics[width=\textwidth]{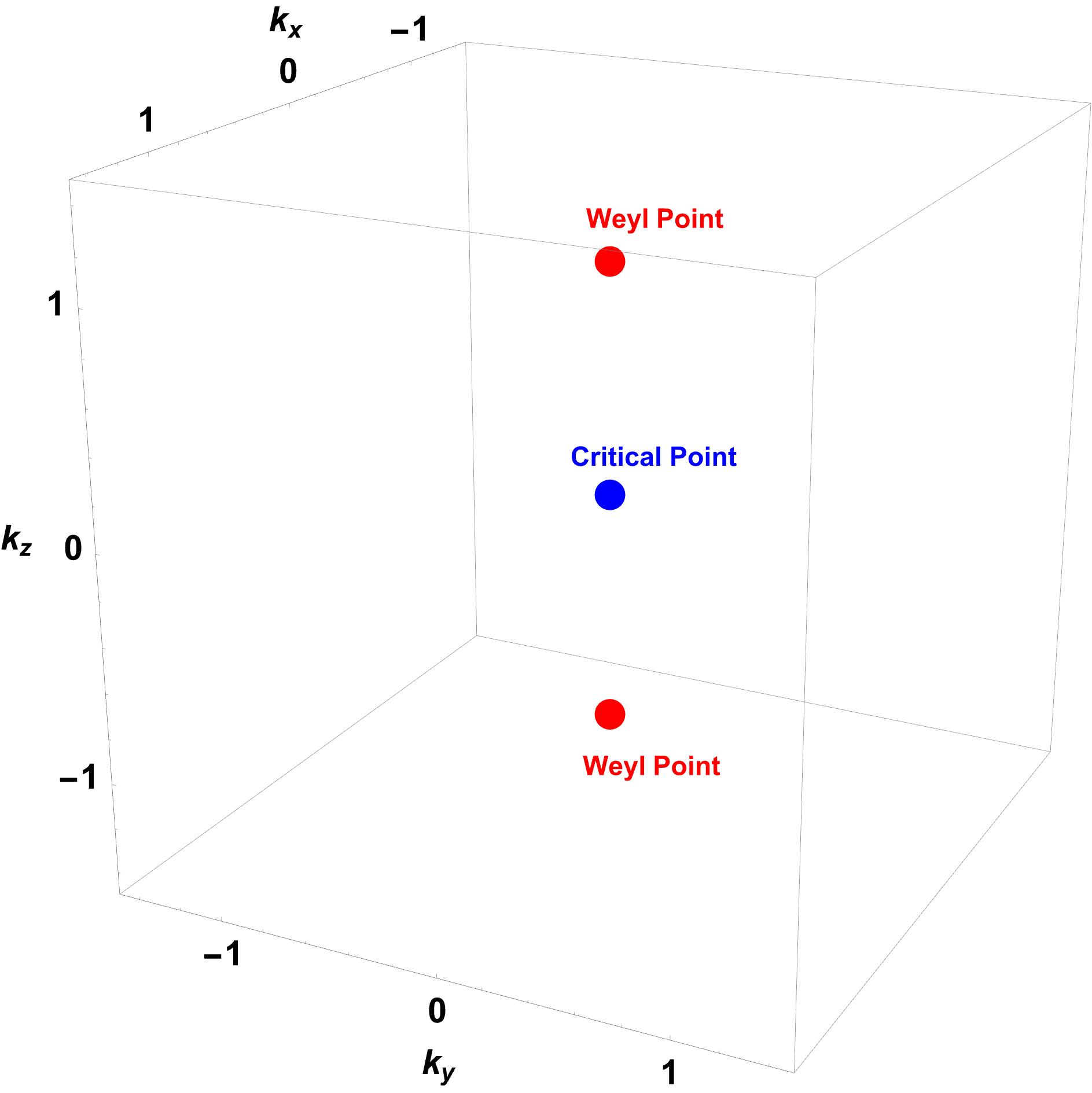}}
            \end{minipage}
            \begin{minipage}{0.49\linewidth}
                \vspace{3pt}
                \centerline{\includegraphics[width=\textwidth]{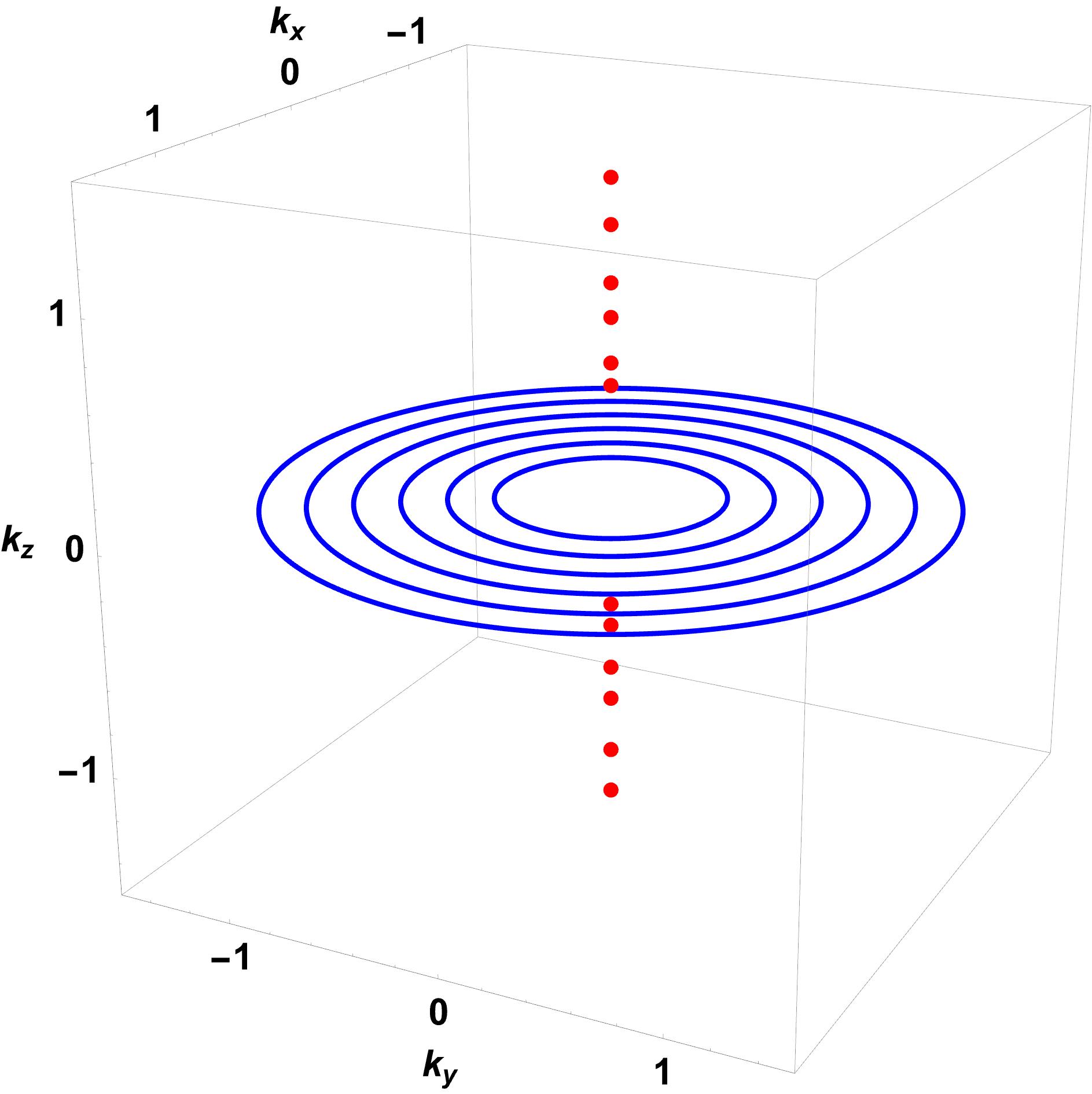}}
            \end{minipage}
            \caption{Left: the Fermi surfaces for weak coupling coexisting semimetals in the Weyl-Critical phase, where the nodal ring shrinks to a critical point. Right:  Fermi surfaces for the Weyl-Critical phase of the holographic coexisting semimetal, where multi Fermi surfaces are presented for both the Weyl sector and the nodal line sector. Weyl nodes are distributed in pairs along the $k_z$ axis as that in the pure holographic Weyl semimetal. There are still infinitely many critical nodal rings in the Weyl-Critical phase. The band structure indicates that the nodal lines at this critical phase are fourfold degenerate. Results of the corresponding topological invariants $\zeta_1,\zeta_2$ through numerical methods also confirm that the topological charges are the same as in the weak coupling case.}
            \label{WeakVsStrongWeylCriticalPhaseBandStructure}
        \end{figure}

        Finally, we have a remark on the crucial difference between the critical phase of the holographic nodal line semimetal or similarly the Weyl-Critical phase in the holographic coexisting semimetal with the corresponding weakly coupled models. {
        At criticality, conventional weak-coupling theory predicts that the nodal line should collapse into a critical point at the critical phase.
        However, our topological Hamiltonian band calculations for the holographic nodal line semimetal reveal that the nodal line does not fully contract to a point at the critical phase as shown in the right figure of Fig. \ref{WeakVsStrongWeylCriticalPhaseBandStructure}.
        Instead, two nodal rings merge into a single nodal ring structure at criticality. Thus there are still infinitely many discrete nodal rings at the critical phase. 
        Each merged nodal ring exhibits a vanishing $\zeta_1$ and $\zeta_2$ invariant, consistent with the critical phase requirements, as a result the nodal ring will be gapped at the topological trivial phase.}

        The  methods and procedures for the Critical-Nodal phase are similar to the Weyl-Critical phase while are numerically much more challenging, so we will not go into details for that case.
\section{Discussion and Outlook}
    \label{Section5}
    There are three topological invariants (Weyl charge/$\zeta_1$ invariant/$\zeta_2$ invariant) in the Weyl-nodal line coexisting semimetal. A nonzero Weyl charge indicates that the Weyl node is topological protected. A non zero $\zeta_1$ invariant indicates that the nodal line is topological protected, while a nonzero $\zeta_2$ invariant denotes that after the nodal line shrinks to a point it will be re-expanded when tuning the order parameter monotonically. 
    In previous work we have calculated the Weyl charge for the strongly coupled holographic Weyl semimetal, and the $\zeta_1$ invariant in the holographic nodal line semimetal, both in the topologically nontrivial phases. 
    
    In this work, we have systematically computed the topological invariants for the Weyl-nodal line coexisting semimetal, especially we investigate the higher order topological invariant $\zeta_2$ in holographic nodal line semimetal.
     The relationship between the  values of the topological invariant and the corresponding phases is summarized in Table \ref{tab:phase_classification}.
     \begin{table}[htbp]
        \centering
        \begin{tabular}{cccc}
            \toprule
             & $\zeta_1/\zeta_0=1,\zeta_2/\widetilde{\zeta}_2=0$ & $\zeta_1/\zeta_0=0,\zeta_2/\widetilde{\zeta}_2=0$ & $\zeta_1/\zeta_0=0,\zeta_2/\widetilde{\zeta}_2=0$ \\
            \midrule
            Weyl Charge $=\pm 1$ & Weyl-Nodal & Weyl-Critical & Weyl-Gap \\
            Weyl Charge $=0$ & Critical-Nodal & Critical-Critical & Critical-Gap \\
            Weyl Charge $=0$ & Gap-Nodal & Gap-Critical & Gap-Gap \\
            \bottomrule
        \end{tabular}
        \caption{The correspondence between topological phases and their respective invariants in Weyl–nodal line coexisting semimetals—across both strong and weak coupling regimes—is summarized in this table. The data reveal that topological invariants alone cannot differentiate critical phases from gapped ones, necessitating an alternative approach based on entanglement entropy to distinguish between them, which we will discuss in future work.}
        \label{tab:phase_classification}
    \end{table}

    We also carefully study the effective band structures and topological invariants in the critical phases in holographic semimetals, including the holographic Weyl, nodal line and coexisting semimetals. These results reveal the following important and unique features for strongly coupled topological semimetals using the tools of AdS/CFT correspondence. First, a  band crossing ordering interchange phenomenon has been observed in holographic nodal line and coexist systems, in parallel to  the behavior previously found in the holographic Weyl semimetal.
    Second, multiple Fermi surfaces exist at the critical phases for all holographic semimetal states, where two Weyl nodes or two nodal rings merge to form one critical nodal point or ring at the critical phase, further confirmed by the calculation of topological invariants in each case. 
    Finally, strikingly different band structures  between the critical phases of the holographic Weyl and nodal line semimetals have been found, which could be attributed to their distinct IR 
   equations for bulk probe fermions.

    Several open questions are in order. First it should be noted that the type of coexisting topological semimetals  in the present work is obtained by introducing an eight-component spinor in the weakly coupled  field theoretic model \eqref{PureWeylLagrangian}. 
    Thus in the corresponding holographic model \eqref{CoexistHolographicModel}, it is necessary to utilize two sets of scalar fields and two eight-component spinors.
    However, the coexisting topological semimetal and a more complicated phase diagram can also be realized by employing only one four-component spinor in the weakly coupled effective field theory, which leads to more novel critical phases such as a triple degenerate nodal point and a critical state with three nodes\cite{TopologicalPhaseTransitionJi:2023rua}. 
    These novel critical phases are considered to be of vital importance in the ``material universe"\cite{PhysRevB.94.165201} and therefore merit detailed study of their properties in the strong coupling regime. 
    Moreover, the corresponding geometries that correspond to these states in the holographic model would also serve as a valuable addition to the holographic dictionary. We hope to report our work in this direction in the future. 
    
    Second, entanglement entropy can be regarded as order parameters for topological phase transitions. This has been studied for a holographic nodal-line semimetal in \cite{Baggioli:2023ynu}. An interesting future direction would be to analyze the entanglement structures of holographic semimetals to reveal the relation between quantum entanglement properties and the topological structure in strongly coupled holographic semimetal systems, especially more refined entanglement structures could be studied by analyzing more smaller subsystems as in \cite{Ju:2024hba, Ju:2024kuc}.
    
    Thirdly, although the topological phases are considered to be robust against environment-induced decoherence, it remains an open question how to determine the topological robustness of a system in conditions that deviate from equilibrium. The Schwinger-Keldysh (SK) effective action is a powerful tool for investigating non-equilibrium systems\cite{Crossley:2015evo}. Its use in chiral fluids\cite{Baggioli:2024zfq} and strange metals\cite{Liu:2024tqe} has been demonstrated to be highly effective. The combination of the SK effective theories with the holographic model is an intriguing avenue for exploration, as it facilitates the investigation of the impact of dissipations and noises on the robustness of a topological system.

\subsection*{Acknowledgments}
We thank K. Landsteiner and Y. Liu for helpful discussions. This work was supported by the National Natural Science Foundation of China (Grant Nos.12035016, 12405078).
\appendix
    \section{Spin connection in curved space}
        \label{AppendixSpinConnection}
        The following section provides the definition of the spin connection for a curved space. Without loss of generality, on a manifold $M$ equipped with a metric $g$, it is possible to choose an orthonormal frame $e_i$ and its dual coframe $\{\theta^i\}$ . Then the connection 1-form ${\omega_j}^i$ defines the covariant derivative as
        \begin{equation}
            D e_i = {\omega_i}^j \otimes e_j.
        \end{equation}
    
        The corresponding connection form, denoted by the symbol $\omega$, which is defined on the frame bundle $F(M)$, relates to the connection 1-form on the base manifold $M$ via the pullback by a section, denoted by the $\sigma: M \to F(M)$:
        \begin{equation}
            \sigma^*\omega := \begin{pmatrix}
                {\omega_1}^1 & \cdots & {\omega_n}^1 \\
                \vdots & \ddots & \vdots \\
                {\omega_1}^n & \cdots & {\omega_n}^n
            \end{pmatrix}: T_p M \to \mathfrak{o}(1,n),
        \end{equation}
        where $\sigma(p) = (p, e_1, \dots, e_n)$. If the Stiefel-Whitney classes $w_1(M)$ and $w_2(M)$ vanish, the frame bundle $F(M)$ reduces to a spin structure $Sp(M)$ with structure group $Spin(1,n)$ and homomorphism $\rho: Spin(1,n) \to SO(1,n)$.
    
        The local connection form $\hat{\omega}$ on $Sp(M)$ is induced by $\omega$ via $\hat{\omega} = (\rho_*)^{-1} \circ \sigma^* \omega$. For the Levi-Civita connection (with ${\omega_j}^i = -{\omega_i}^j$), this simplifies to
        \begin{equation}
            \hat{\omega} = \frac{1}{2} \sum_{i<j} {\omega_i}^j e_i e_j: T_p M \to \mathfrak{spin}(1,n),
        \end{equation}
        where $e_i e_j \in Cl(T_p M)$ is the Clifford product.

        The spinor bundle is built from the $Spin(1,n)$-representation $\tau:Spin(1,n)\to GL(\Psi)$. Assuming that the $Sp(M)$ is a spin structure on $M$ and employing the representation $\tau$, the associated vector bundle $S(M):=Sp(M)\times_\tau\Psi$ is defined by quotienting $Sp(M)\times\Psi$ under the equivalence relation $(p\cdot g,v)\sim(p,\tau(g)v)$ for $g\in Spin(1,n)$. $S(M)$ is a vector bundle with fiber $\Psi$, whose sections are spinor fields.
            
        The covariant derivative of a spinor field $\Psi$ is given by\cite{SpinConnectionlawson2016spin}:
        \begin{equation}
            D\Psi=d\Psi+\hat{\omega}\cdot\psi=d\Psi+\frac{1}{4}\sum_{ij} {\omega_i}^j (e_i\cdot e_j-e_j\cdot e_i)\cdot\Psi,
        \end{equation}
        where $\cdot$ denotes the Clifford product. In terms of gamma matrices, this can be rewritten in a more conventional form by defining $\Gamma^a:=\eta^{ab}e_b\cdot$ in the Lorentzian manifold:
        \begin{equation}
            D\Psi=d\Psi+\frac{1}{4}\sum_{ab} {\omega_a}^b[\Gamma^a,\Gamma^b]\Psi.
        \end{equation}
        Here, $\eta = {\rm diag\ }(1,-1,-1,-1,-1)$ is the metric signature, and the spinor space inner product is defined as $(\Psi_1,\Psi_2):=i\Psi_1^\dag\Gamma^0\Psi_2$. This formulation satisfies the Clifford algebra relations:
        \begin{equation}
            \{\Gamma^a,\Gamma^b\}=-2\eta^{ab}, \quad (\Psi_1,\Gamma^a\Psi_2)=-(\Gamma^a\Psi_1,\Psi_2)
        \end{equation}
        The Dirac operator is given by $\Gamma^a D_{e_a}$, which leads to the equation of motion for a free fermion of mass $m\in\mathbb{R}$:
        \begin{equation}
            \Gamma^aD_{e_a}\Psi-m\Psi=0.
        \end{equation}

        In particular, when the metric takes the simple diagonal form
        \begin{equation}
            g = \sum_a \eta^{aa} g_{aa} dx^a \otimes dx^a,
        \end{equation}
        the connection 1-form can be expressed explicitly. First, we define the orthonormal frame and its dual coframe:
        \begin{equation}
            \theta^a = \sqrt{g_{aa}} dx^a, \quad e_a = \frac{1}{\sqrt{g_{aa}}} \frac{\partial}{\partial x^a}.
        \end{equation}
        Using Cartan's structure equation $d\theta^a = \theta^b \wedge \omega_b{}^a$, we obtain the explicit expression for the connection 1-form $\omega_b{}^a$:
        \begin{equation}
            \begin{aligned}
                {\omega_b}^a&=\frac{1}{\sqrt{g_{bb}}}\frac{\partial\sqrt{g_{aa}}}{\partial x^b}dx^a+\frac{1}{\sqrt{g_{aa}}}\frac{\partial\sqrt{g_{bb}}}{\partial x^a}dx^b,~~a=0,b\ne 0 ~or~a\ne 0,b=0,\\
                {\omega_b}^a&=\frac{1}{\sqrt{g_{bb}}}\frac{\partial\sqrt{g_{aa}}}{\partial x^b}dx^a-\frac{1}{\sqrt{g_{aa}}}\frac{\partial\sqrt{g_{bb}}}{\partial x^a}dx^b,~~else.
            \end{aligned}
        \end{equation}
    
        The covariant derivative can be written explicitly
        \begin{equation}
            D_{e_c}\Psi=\frac{1}{\sqrt{g_{cc}}}\frac{\partial\Psi}{\partial x^c}+\frac{1}{4}\sum_{ab}\braket{{\omega_a}^b,e_c}[\Gamma^a,\Gamma^b]\Psi.
        \end{equation}
        
        Without loss of generality, we choose the Gamma matrices as chiral-Weyl representation:
        \begin{equation}
            \begin{aligned}
                I_2&=\begin{pmatrix}1 & 0 \\ 0 & 1\end{pmatrix},~\sigma^1=\begin{pmatrix}0 & 1 \\ 1 & 0\end{pmatrix},~\sigma^2=\begin{pmatrix}0 & -i \\ i & 0\end{pmatrix},~\sigma^3=\begin{pmatrix}1 & 0 \\ 0 & -1\end{pmatrix},\\
                \gamma^0&=\begin{pmatrix}0 & I_2 \\ I_2 & 0\end{pmatrix},~\gamma^1=\begin{pmatrix}0 & \sigma^1 \\ -\sigma^1 & 0\end{pmatrix},~\gamma^2=\begin{pmatrix}0 & \sigma^2 \\ -\sigma^2 & 0\end{pmatrix},~\gamma^3=\begin{pmatrix}0 & \sigma^3 \\ -\sigma^3 & 0\end{pmatrix},\\
                \gamma^5&=i\gamma^0\gamma^1\gamma^2\gamma^3=\begin{pmatrix}-I_2 & 0 \\ 0 & I_2         \end{pmatrix},~(\Gamma^t,\Gamma^x,\Gamma^y,\Gamma^z,\Gamma^r)=(i\gamma^0,i\gamma^1,i\gamma^2,i\gamma^3,-\gamma^5).
            \end{aligned}
        \end{equation}
        
        In this work, we adopt the abbreviated notation: $D_a \equiv D_{e_a}$ for the covariant derivative along the frame field $e_a$.
        
    \section{IR solution for probe fermions}
        \label{ProbeFermionNearHorizonSolution}
        The subsequent section provides a comprehensive solution for the probe femions employed in the holographic coexisting Weyl nodal line semimetal case under the IR geometry \eqref{CoexistWeylNodalIRGeometry}.
        
        The equation of motion for a free fermion with mass m can be given as:  
        \begin{equation}
            (\Gamma^aD_a-m)\psi=0,
        \end{equation}
        where the Gamma matrix $\Gamma^a$ satisfies with $\{\Gamma^a,\Gamma^b\}=-2\eta^{ab}$. $\eta$ matrix can be defined as $\eta\equiv{\rm diag}(1,-1,-1,-1,-1)$. 
        
        In the following part, the infrared solutions of probe fermions will be demonstrated in two independent cases: the holographic nodal line semimetal and the Weyl semimetal. Subsequently, the solution for the probe fermions in the holographic coexisting Weyl semimetal is obtained by taking the direct sum of these two independent cases.

        In the case of the holographic nodal line semimetal system, the leading order for the equation of motion near the horizon is
        \begin{equation}
            \label{NodalHorizonEquation}
            \left(\Gamma^r\frac{\partial}{\partial r}+\frac{ik_\mu\Gamma^\mu}{r^2}\right)\psi=0,
        \end{equation}
        where $k_\mu=\left(-\frac{\omega}{u_0}, 0, 0, \frac{k_z}{u_0}\right),~k^2=k_\mu k_\nu\eta^{\mu\nu}=(i k_\mu\Gamma^\mu)(i k_\nu\Gamma^\nu)$.
        Apply $\Gamma_\pm$ to both sides of the equation \eqref{NodalHorizonEquation} we have
        \begin{equation}
            \pm\frac{\partial}{\partial r}\psi_\pm+\frac{ik_\mu\Gamma^\mu}{r^2}\psi_\mp=0,
        \end{equation}
        with the notation $\Gamma_\pm:=\frac{1\pm\Gamma^r}{2},~\Gamma_\pm\psi=\psi_\pm$. Then it is possible to divide the equation of motion into two uncoupled parts
        \begin{equation}
            \label{SeperatedNodalHorizonEquation}
            \psi_\pm=\pm\frac{i k_\mu\Gamma^\mu}{k^2}r^2\frac{\partial}{\partial r}\psi_\mp,~
            \psi_\pm(r)=-\left(\frac{r^2}{k}\frac{\partial}{\partial r}\right)^2\psi_\pm(r).
        \end{equation}

        For the sake of simplicity, it is permissible to substitute the variable $r$ for $r=\frac{k}{z}$. Subsequently, the \eqref{SeperatedNodalHorizonEquation} is rendered as follows:
        \begin{equation}
            \psi_\pm(z)=-\left(-\frac{\partial}{\partial z}\right)^2\psi_\pm(z).
        \end{equation}
        
        Take $\psi_+$ as an example, the solution is easy to get
        \begin{equation}
            \psi_+=C_1e^{i\frac{k}{r}}+C_2e^{-i\frac{k}{r}},~\psi_-=-\frac{k_\mu\Gamma^\mu}{k}C_1 e^{i\frac{k}{r}}+\frac{k_\mu\Gamma^\mu}{k}C_2 e^{-i\frac{k}{r}}.
        \end{equation}
        After taking the in-going horizon condition, there are only one branch of solution is reasonable
        \begin{equation}
            \psi_+=Ce^{i\frac{k}{r}},~\psi_-=-\frac{k_\mu\Gamma^\mu}{k}Ce^{i\frac{k}{r}}.
        \end{equation}
        So the near horizon solution for probe fermion in holographic nodal line semimetal is
        \begin{equation}
            \psi=e^{i\frac{k}{r}}\left(1-\frac{k_\mu\Gamma^\mu}{k}\right)C,~C\in\Im\Gamma_+,
        \end{equation}
        or
        \begin{equation}
            \psi=e^{i\frac{k}{r}}\left(1+\frac{k_\mu\Gamma^\mu}{k}\right)C,~C\in\Im\Gamma_-.
        \end{equation}
        
        For the case of the holographic Weyl semimetal, the near horizon equation of motion is 
        \begin{equation}
            \label{WeylHorizonEquation}
            \left(\Gamma^r\frac{\partial}{\partial r}+\frac{ik_\mu\Gamma^\mu}{r^2}-\frac{m}{r}\right)\psi=0,
        \end{equation}
        where $k_\mu=(-\omega,k_x,k_y,k_z-q a_0), k^2=k_\mu k_\nu\eta^{\mu\nu}=(i k_\mu\Gamma^\mu)(i k_\nu\Gamma^\nu)$.
        
        Similar, by applying $\Gamma_\pm$ to both sides of the equation \eqref{WeylHorizonEquation} we have
        \begin{equation}
            \pm\frac{\partial}{\partial r}\psi_\pm+\frac{ik_\mu\Gamma^\mu}{r^2}\psi_\mp-\frac{m}{r}\psi_\pm=0,
        \end{equation}
        with the notation $\Gamma_\pm:=\frac{1\pm\Gamma^r}{2},~\Gamma_\pm\psi=\psi_\pm$. Then it is possible to divide the equation of motion into two uncoupled parts
        \begin{equation}
            \label{SeperatedWeylHorizonEquation}
            \psi_\mp=\frac{ik_\mu\Gamma^\mu}{k^2}r\left(\mp r\frac{\partial}{\partial r}+m\right)\psi_\pm,~\psi_\pm=\frac{1}{k^2}\left(\pm r^2\frac{\partial}{\partial r}+mr\right)\left(\mp r^2\frac{\partial}{\partial r}+mr\right)\psi_\pm.
        \end{equation}

        Changing the variables $r=\frac{k}{z}$, the \eqref{SeperatedWeylHorizonEquation} has a simple form
        \begin{equation}
            \begin{aligned}
                \psi_\pm=\left(\mp \frac{\partial}{\partial z}+\frac{m}{z}\right)\left(\pm \frac{\partial}{\partial z}+\frac{m}{z}\right)\psi_\pm,\\
                \frac{\partial^2}{\partial z^2}\psi_\pm-\frac{m^2}{z^2}\psi_\pm\mp\frac{m}{z^2}\psi_\pm+\psi_\pm=0.
            \end{aligned}
        \end{equation}
        Performing the transformation $\psi_\pm=\sqrt{z}f_\pm$, we can change the equation of motion into a typical Bessel equation
        \begin{equation}
            \frac{d^2f_\pm}{dz^2}+\frac{1}{z}\frac{df_\pm}{dz}+\left(1-\frac{\left(m\pm\frac{1}{2}\right)^2}{z^2}\right)f_\pm=0,
        \end{equation}
        whose solution is known
        \begin{equation}
            \psi_{\pm}=C_1\frac{1}{\sqrt{r}}J_{m\pm\frac{1}{2}}\left(\frac{k}{r}\right)+C_2\frac{1}{\sqrt{r}}Y_{m\pm\frac{1}{2}}\left(\frac{k}{r}\right).
        \end{equation}
        Bessel J and Bessel Y function can be written in a series form
        \begin{equation}
            \label{FirstKindBesselFunction}
            \begin{aligned}
                J_\alpha(x)&=\sum_{k=0}^\infty\frac{(-1)^k}{k!\Gamma(k+\alpha+1)}\left(\frac{x}{2}\right)^{2k+\alpha},\\
                Y_\alpha(x)&=\frac{\cos\pi\alpha J_{\alpha}(x)-J_{-\alpha}(x)}{\sin\pi\alpha}.
            \end{aligned}
        \end{equation}
        
        After taking the in-going horizon condition, there are only one branch of solution is reasonable
        \begin{equation}
            \psi_+=\frac{C}{\sqrt{r}}J_{m+\frac{1}{2}}\left(\frac{k}{r}\right),~\psi_-=\frac{ik^\mu\Gamma^\mu}{k}\frac{C}{\sqrt{r}}J_{m-\frac{1}{2}}\left(\frac{k}{r}\right),~C\in\Im\Gamma_+.
        \end{equation}
        
        It is similar for the probe fermion with negative mass $-m$
        \begin{equation}
            \begin{aligned}
                \frac{d^2f_\pm}{dz^2}+\frac{1}{z}\frac{df_\pm}{dz}+\left(1-\frac{\left(m\mp\frac{1}{2}\right)^2}{z^2}\right)f_\pm=0,~\psi_\pm=\sqrt{z} f_\pm,\\
                \psi_-=\frac{C}{\sqrt{r}}J_{m+\frac{1}{2}}\left(\frac{k}{r}\right),~\psi_+=-\frac{ik^\mu\Gamma^\mu}{k}\frac{C}{\sqrt{r}}J_{m-\frac{1}{2}}\left(\frac{k}{r}\right),~C\in\Im\Gamma_-.
            \end{aligned}
        \end{equation}
        
        It should be noted that the $\omega$ is quite small, which results in $k^2<0$. It is therefore more expedient to utilise the second Kind Bessel equation:
        \begin{equation}
            \begin{aligned}
                \frac{d^2f_\pm}{dz^2}+\frac{1}{z}\frac{df_\pm}{dz}+\left(1+\frac{\left(m\pm\frac{1}{2}\right)^2}{z^2}\right)f_\pm=0,~z=\frac{|k|}{r},~\psi_\pm=\sqrt{z} f_\pm,\\
                \psi_{\pm}=C_1\frac{1}{\sqrt{r}}I_{m\pm\frac{1}{2}}\left(\frac{|k|}{r}\right)+C_2\frac{1}{\sqrt{r}}K_{m\pm\frac{1}{2}}\left(\frac{|k|}{r}\right),
            \end{aligned}
        \end{equation}
        where Bessel I and Bessel K is
        \begin{equation}
            \label{SecondKindBesselFunction}
            \begin{aligned}
                I_\alpha(x)=i^{\alpha} J_\alpha(ix),\\
                K_\alpha(x)=\frac{\pi}{2}\frac{I_{-\alpha}(x)-I_{\alpha}(x)}{\sin\pi\alpha}.
            \end{aligned}
        \end{equation}
        
        Considering the in-going horizon condition, there are only one branch of solution is reasonable
        \begin{equation}
            \psi_+=\frac{C}{\sqrt{r}}K_{m+\frac{1}{2}}\left(\frac{|k|}{r}\right),~\psi_-=\frac{ik_\mu\Gamma^\mu}{|k|}\frac{C}{\sqrt{r}}K_{m-\frac{1}{2}}\left(\frac{|k|}{r}\right),~C\in\Im\Gamma_+.
        \end{equation}

        It is similar for the probe fermion with negative mass $-m$
        \begin{equation}
            \begin{aligned}
                \frac{d^2f_\pm}{dz^2}+\frac{1}{z}\frac{df_\pm}{dz}+\left(1+\frac{\left(m\mp\frac{1}{2}\right)^2}{z^2}\right)f_\pm=0,~z=\frac{|k|}{r},~\psi_\pm=\sqrt{z} f_\pm,\\
                \psi_-=\frac{C}{\sqrt{r}}K_{m+\frac{1}{2}}\left(\frac{|k|}{r}\right),~\psi_+=-\frac{ik_\mu\Gamma^\mu}{|k|}\frac{C}{\sqrt{r}}K_{m-\frac{1}{2}}\left(\frac{|k|}{r}\right),~C\in\Im\Gamma_-.
            \end{aligned}
        \end{equation}
        

\bibliographystyle{elsarticle-num}
\bibliography{biblio}

\end{document}